\documentclass[aps,prb,twocolumn,showpacs,showkeys,superscriptaddress,floatfix]{revtex4}
\usepackage{amsmath,graphicx}
\usepackage{dcolumn}
\usepackage{bm}
\usepackage{xcolor}

\definecolor{mblue}{rgb}{.23,.326,.643}
\definecolor{mgreen}{rgb}{.251,.5,.173}
\definecolor{mred}{rgb}{.93,.133,.141}
\definecolor{mmagenta}{rgb}{.725,.325,.621}
\begin{document}
\title{Strain and band-mixing effects on the excitonic Aharonov-Bohm effect in In(Ga)As/GaAs ringlike quantum dots}

\author{Vladimir V. Arsoski}\email{vladimir.arsoski@etf.bg.ac.rs}
\affiliation{School of Electrical Engineering, University of
Belgrade, P.O. Box 35-54, 11120 Belgrade, Serbia}

\author{Milan \v{Z}. Tadi\'c}\email{milan.tadic@etf.bg.ac.rs}
\affiliation{School of Electrical Engineering, University of
Belgrade, P.O. Box 35-54, 11120 Belgrade, Serbia}

\author{Fran\c{c}ois M. Peeters}\email{francois.peeters@ua.ac.be}
\affiliation{Department of Physics, University of Antwerp,
Groenenborgerlaan 171, B-2020 Antwerp, Belgium}

\date{\today}
\begin{abstract}
Neutral excitons in strained axially symmetric In(Ga)As/GaAs quantum
dots with ringlike shape are investigated. Similar to experimental
self-assembled quantum rings, the analyzed quantum dots have
volcano-like shapes. The continuum mechanical model is employed to
determine the strain distribution, and the single-band envelope function
approach is adopted to compute the electron states. The hole states are
determined by the axially symmetric multiband Luttinger-Kohn
Hamiltonian, and the exciton states are obtained from an exact
diagonalization. We found that the presence of the inner layer
covering the ring opening enhances the excitonic Aharonov-Bohm (AB)
oscillations. The reason is that the hole becomes mainly localized in
the inner part of the quantum dot due to strain, whereas the electron
resides mainly inside the ring-shaped rim. Interestingly, larger AB
oscillations are found in the analyzed quantum dot than in a fully
opened quantum ring of the same width. Comparison with the unstrained
ring-like quantum dot shows that the amplitude of the excitonic Aharonov-Bohm
oscillations are almost doubled in the presence of strain. The computed oscillations
of the exciton energy levels are comparable in magnitude to the oscillations
measured in recent experiments.
\end{abstract}



\maketitle

\section{Introduction}

Semiconductor quantum dots (QD) have been widely investigated in the
last decade because of their potential applications for electronic\cite{Kastner2000},
photonic\cite{Michler2000}, spintronic\cite{Engel2004}, and quantum
computing devices\cite{Engel2004,Engel2001}. Different fabrication procedures are adopted to grow these structures, and quantum dots of various shapes and
dimensions have been realized. A peculiar example is a quantum ring
(QR)\cite{Garcia1997,Teodoro2012}, which has a doubly connected topology\cite{Lee2004}. Hence, the Aharonov-Bohm (AB) effect\cite{AB1959}, which arises from a change of phase of the particle wave function with magnetic field, takes place in the scattering-free limit. In circular ring-like structures the AB effect is manifested by the crossings of energy levels of different angular momenta in the ground state, which are referred to as \textit{angular momentum transitions}, and which lead to oscillations of the
electron ground energy level with magnetic field\cite{Lee2004}.

In addition to the AB effect for single particle states, the ground
energy level of the neutral exciton may exhibit oscillations when the
external magnetic field through the QR varies. This \textit{excitonic
AB effect} was predicted for concentric one-dimensional (1D) rings\cite{Govorov2002},
 where the electron and hole are found to accumulate different
phases in each revolution in the presence of an external magnetic field.
Also, this simple model predicted that the bright exciton state can be
turned into a dark one. These bright-to-dark exciton transitions produce
peculiar oscillations of the emission intensity with magnetic field,
which are referred to as the \textit{optical excitonic} \textit{AB}
\textit{effect}\cite{Govorov2002}. In a three-dimensional (3D) geometry, however, the overlap
between the electron and hole wave functions are finite, thus the
emission intensity does not exhibit periodic oscillations as in 1D
quantum rings. Rather, due to a small polarization of the 3D exciton,
small oscillations of the exciton ground energy level, which are of the
order of a fraction of meV, were experimentally observed\cite{Ding2010,Teodoro2010}.

Most works on the excitonic AB effect has been focused on In(Ga)As/GaAs
quantum rings, which are self-assembled by means of the
Stranski-Krastanov (SK) growth,\cite{Engel2001} and are strained because of
lattice mismatch between In(Ga)As and GaAs. Strain is involved in both
the initial phase of the self-assembly of lens-shaped quantum dots, and
the subsequent dewetting procedure which leads to an outward diffusion
of In adatoms from the central part of the InAs islands. As a result,
the formed quantum rings have volcano-like shapes, and exhibit
substantial compositional intermixing between the dot and the matrix
\cite{Fomin2007}. The redistribution process has been demonstrated to be
strongly temperature dependent\cite{Lorke2002}, and QR's of various
composition and morphology have been fabricated\cite{Ding2010,Teodoro2010,
Fomin2007,Lorke2002,Granados2003,Kleemans2007,Offermans2005}.
For example, self-assembled quantum rings exhibiting considerable in-plane
anisotropy have been explored in Refs.~\onlinecite{Granados2003}--\onlinecite{Offermans2005}.
More recently, almost axially symmetric In(Ga)As/GaAs quantum rings were fabricated and
analyzed\cite{Teodoro2010}, but no height versus radius dependence of these rings
was determined.

When the QR opening is covered by the material of the ring, the AB
oscillations of the single-particle states are considerably modified
\cite{Fomin2007,Kleemans2007,Arsoski2012,Li2011,Cukaric2012}.
In view of the recent experimental discovery of the excitonic AB effect
in type-I quantum rings\cite{Ding2010,Teodoro2010}, a few issues deserve special
attention. \textit{First}, increasing the height of the QD rim with
respect to the height of the inner layer [see Fig.\:1] enables larger
space for both the electron and hole in the rim, and therefore the
exciton states could exhibit a QR-like behavior\cite{Granados2003,Arsoski2012,Li2011,Cukaric2012,Tadic2011,Filikhin2006}.
\textit{Second}, band mixing could have an important impact on the
hole states, and in turn their dependence of the exciton ground state energy on the
magnetic field\cite{Tadic2011,Climente2003}. \textit{Third}, recent theoretical
results have indicated that the spatial variation of strain in fully
opened quantum rings is beneficial for the appearance and the magnitude
of excitonic AB oscillations\cite{Arsoski2012,Tadic2011}. Strain could lead to an
effective separation of the electron and hole, thereby increasing the
exciton polarization. A similar separation of the electron and hole in
type-II quantum dots was previously found to promote excitonic AB
oscillations\cite{Ribeiro2004}.

In this paper, we explore how the layer covering the quantum ring
opening affects the excitonic Aharonov-Bohm effect. We assume that the
analyzed ringlike quantum dot (RLQD) is axially symmetric, as depicted in Fig.\:1(a).
The magnetic field dependence of the electron, hole, and exciton
states are computed for a few values of the rim height and fixed
height of the inner layer, as illustrated in Fig.\:1(b). The strain
distribution is obtained within the approximation of isotropic elasticity
in the continuum mechanical model\cite{Davies1998,Tadic2002}.
The electron envelope functions are computed within the single-band effective
mass approach, whereas the hole envelope functions are extracted from both the axial and
spherical approximations of the multiband Luttinger-Kohn model. The
exciton states are calculated using an exact numerical diagonalization. We
will investigate how the size of the rim affects the excitonic AB
effect. Furthermore, we explore effects of band mixing on the magnetic
field dependence of the exciton energy levels. We especially analyze how
strain affects the mixing of the heavy-hole (HH) and light-hole (LH)
states and what are the consequences of this mixing on the exciton
states.

The paper is organized as follows. In Sec. II, we present the
theoretical framework to compute the strain distribution, the
single-particle, and the exciton states. The results of our numerical
calculations are presented and discussed in Sec. III. Our conclusions
are given in Sec. IV.

\begin{figure}
     \begin{center}
       \includegraphics[width=8.6cm]{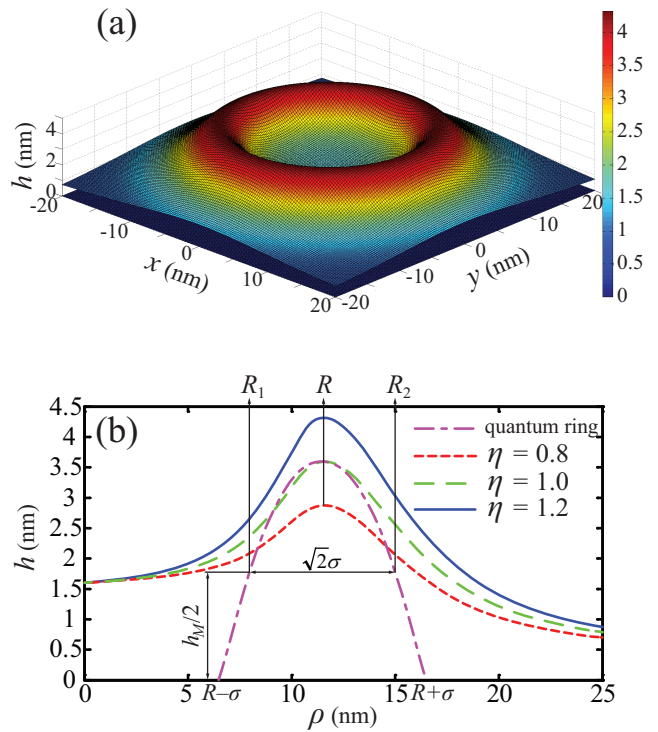}
	\caption{\label{fig1}(Color online.) (a) Profile of the analyzed ringlike quantum dot
	(RLQD). (b) The radial dependence of the RLQD height \textit{h} for three
	values of the parameter {\it $\eta$} that appears in Eq.\:(5). The cross section
	of the quantum ring defined by Eq.\:(6) is shown by the dash-dotted line.}
    \end{center}
\end{figure}

\section{Theoretical models}

\subsection{Model of mechanical strain}

The analyzed RLQD is composed of In(Ga)As, which is embedded in an infinite
GaAs matrix, and is strained due to the lattice mismatch between
In(Ga)As and GaAs. An important theoretical model that allows one to
calculate strain was introduced by Dawnes\cite{Davies1998}, who based his
derivations on the Eshelby inclusion theory\cite{Eshelby1957}. The model was
subsequently applied to quantum dots of different shape and composition
\cite{Davies1998,Downes1997}. In this approach, the elastic behavior of the material is
assumed to be linear and isotropic. Also, the dot is assumed to be
embedded in an infinite matrix, with a lattice constant mismatch
\begin{equation}
	\varepsilon_{0}=\frac{({\it a_{d}}-{\it a_{m}})}{\it a_{m}},
\end{equation}
between the dot and the matrix. Here, ${\it a_{m}}$ denotes the
lattice constant of the matrix material, and ${\it a_{d}}$ is
the lattice constant in the dot. The following equation is obtained for
axially symmetric quantum dots
\begin{equation}
	\frac{1}{\rho}\frac{\partial}{\partial \rho}\left(\rho \frac{\partial \chi}{\partial \rho}\right)+\frac{\partial ^2 \chi}{\partial z^2}=
	\frac{1+\nu}{1-\nu}\varepsilon_{\it d}(\bf{r}).
\end{equation}
Here, {$\varepsilon_{d}(\bf r)$} is equal to {$\varepsilon_{0}$} in the dot and is zero in the matrix, $\chi$ denotes the Lam\'e displacement potential and
{$\nu$} is the Poisson ratio. The displacement vector {${\bf u}={\bf grad}\chi=\frac{\partial\chi}{\partial\rho}{\bf e}_{\rho}+\frac{\partial\chi}{\partial z}{\bf e}_{z}$}
 and the components of the strain tensor
\begin{subequations}
	\begin{equation}
	\varepsilon_{\rho \rho}=\frac{\partial u_{\rho}}{\partial \rho},
    \end{equation}
	\begin{equation}    	
	\varepsilon_{\phi \phi}=\frac{u_{\rho}}{\rho},
    \end{equation}
	\begin{equation}
	\varepsilon_{zz}=\frac{\partial u_{z}}{\partial z},
    \end{equation}
	\begin{equation}
	\varepsilon_{\phi z}=\varepsilon_{\rho \phi}=0,
    \end{equation}
	\begin{equation}
	\varepsilon_{\rho z}=\frac{1}{2}\left(\frac{\partial u_{\rho}}{\partial z}+\frac{\partial u_{z}}{\partial \rho}\right),
    \end{equation}
\end{subequations}
are determined from the computed $\chi=\chi(\rho,\phi,z)$, which is axially symmetric in the analyzed RLQD. To obtain the total solution for the components of the strain tensor from Eqs.\:(2) and (3), the term {$-\varepsilon_{d}({\bf r})$}, that accounts for the initial compression, must be added to the tensile strains, while no correction is needed for the shear strain components. Furthermore, the hydrostatic strain in an arbitrary
shaped quantum dot is given by \cite{Davies1998}
\begin{equation}
	\varepsilon_{hyd}=\varepsilon_{\rho \rho}+\varepsilon_{\phi \phi}+\varepsilon_{zz}=-2\frac{1-2\nu}{1-\nu}\varepsilon_{d}({\bf r}),
\end{equation}
and therefore, for a homogenous material it is constant inside the dot
and equals zero in the matrix. To avoid the division by $\rho$ in Eq.~(3b), {$\varepsilon_{\phi \phi}$} is calculated by using Eq.\:(4).

\subsection{Electron states in the conduction band}

Our aim is to analyze nearly cylindrically symmetric quantum rings which
were experimentally explored in Ref.~\onlinecite{Teodoro2010}. But their dimensions were not accurately determined, and therefore we used the cross sections of the quantum rings explored in Refs.~\onlinecite{Fomin2007} and \onlinecite{Offermans2005} in the planes $(1\overline{1}0)$, $(100)$, and $(110)$ [see Fig.\:1(b)], and revolved them around the $[001]$ direction to produce axially symmetric shapes. The volcano-like shape of such a formed RLQD is shown in Fig.\:1(a), and following Ref.~\onlinecite{Fomin2007} it is described by the function
\begin{equation}
	h(\rho)=\begin{cases}
	h_0+\frac{(\eta h_M-h_0) \left[1-\left(\rho /R-1\right)^2 \right]}{\left[\left(\rho-R\right)/\gamma_{0} \right]^2+1};&\rho\le R\\
	h_{\infty}+\frac{\eta h_M-h_{\infty}}{\left[\left(\rho-R\right)/\gamma_{\infty} \right]^2+1};&\rho > R
	\end{cases}.
\end{equation}
Here, {\it R} is the radial position of the rim top, {\it h}$_{0}$ is the
height of the RLQD at {\it $\rho$} =0, {\it $h_{\infty}$} is the wetting layer
thickness, {\it $h_{M}$} is the reference value for the height of the RLQD rim,
whereas the parameters {$\gamma_{0}$} and {$\gamma_{\infty}$} determine
the inner and outer slope of the rim, respectively. The dimensionless
parameter {$\eta$} is the ratio of the RLQD height and {\it $h_{M}$}. For
{$\eta$}=0.8, {$\eta$}=1.0, and {$\eta$}=1.2 the variation of {\it $h$} with
{$\eta$} corresponds to the  (1$\overline{1}$0), (100), and (110)
 cross section of the ring of Ref.~\onlinecite{Fomin2007}, respectively.
The values of the parameters used to generate the curves in Fig.\:1(b) are:
{\it $R$} = 11.5 nm, {\it $h_{0}$}= 1.6 nm, {\it $h_{\infty}$} = 0.4 nm,
{\it $h_{M}$} = 3.6 nm, {$\gamma_{0}$} = 3 nm, and  {$\gamma_{\infty}$}= 5 nm.
Note that the analyzed RLQD is assumed to be axially symmetric, which is justified by the fact that the very existence of the AB effect is related to the topology of the structure. The deviations due to the in-plane anisotropy of the structure shape may only affect the period of the AB oscillations\cite{Kleemans2007}. Furthermore, the difference of the rim height for three cross sections shown in Fig.~1(b) is small, and therefore one might infer that the AB oscillations of the single-particle energy spectra for the three values of $\eta$ exhibit small mutual differences. Furthermore, the advantage of adopting the axially symmetric model is its conceptual simplicity.

We will compare the electronic structure of an RLQD with the one of a fully opened ring, whose cross section is described by
\begin{equation}
	h(\rho)=\begin{cases}
	0;&\rho<R-\sigma\\
	h_M\left[1-\left(\frac{\rho-R}{\sigma}\right)^2 \right];&R-\sigma<\rho< R+\sigma\\
	0;&R+\sigma<\rho
	\end{cases},
\label{ring:open}
\end{equation}
where {$\sigma$} is proportional to the full-width at half-maximum [see
dash-dotted curve in Fig.\:1(b)].

The electron states are extracted from the single-band effective-mass
Hamiltonian
\begin{equation}
	\hat{H}_{e}=({\bf p}+q{\bf A}_{e})\frac{1}{2m_{e}^*({\bf r}_{e})}({\bf p}+q{\bf A}_{e})+\hat{H}_{Ze}+V_{e}({\bf r}_{e}).
\end{equation}
Here, {$\it m_{e}^*$} is the electron effective mass, {\it q} is the elementary charge, {\bf p}
 is the canonical momentum operator, and {${\bf A}_{\it e}$} is the magnetic
vector potential which in the Coulomb symmetric gauge has the form
{${\bf A}_{\it e}={\bf B}\times{\bf r}_{\it e}/2$}. Here, {\bf B}
denotes the external magnetic field, which is assumed to be uniform and
directed along the \textit{z} axis. The Zeeman term \textit{$\hat{H}_{Z e}$} has the form
\begin{equation}
	\hat{H}_{Ze}=g_{e}^* \mu_{B}{\bf B}\hat{{\boldsymbol \sigma}},
\end{equation}
where {$\mu_{B}=q\hbar/(2m_{0})$} is the Bohr magneton, {\it m}$_{0}$ is
the free-electron mass, {$g_{e}^*$} is the effective Land\'e {$\it g$}-factor, and
$\hat{{\boldsymbol \sigma}}$  is the Pauli spin matrix. The effective confinement
potential \textit{V$_{e}$} is given by \cite{Tadic2002}
\begin{equation}
    \begin{split}
	V_e({\bf r})&=a_c(\varepsilon_{\rho \rho}+\varepsilon_{\phi \phi}+\varepsilon_{zz})+V_{off,e}({\bf r})\\&=a_c\:\varepsilon_{hyd}({\bf r})+V_{off,e}({\bf r}),
\end{split}
\end{equation}
where {\it a$_{c}$} and {\it $V_{off,e}$} are the deformation potential and
the confining potential due to the offset in the conduction band, respectively.
Because the hydrostatic strain is piecewise constant in the employed
model of isotropic elasticity, the effective potential of the electron
in the conduction band shifts rigidly inside the dot, and is zero in the
matrix.

Because of axial symmetry, the orbital quantum
number {\it $l_{e}$}, which represents quantization of the {\it z} projection of the
electron orbital momentum {\bf L}, is a good quantum number for
the electron state. Thus the electron envelope function is given by
\begin{equation}
	\varPsi_{n_{e},l_{e}}^e({\bf r}_{e})=\frac{1}{\sqrt{2\pi}}e^{il_{e}\phi_{e}}\psi_{n_{e},l_{e}}(\rho_{e},z_{e}),
\end{equation}
where {$n_{e}$} denotes the principal quantum number of the electron state.

\subsection{Hole states}

The hole states are described by the Luttinger-Kohn (LK) Hamiltonian:\cite{Chuang1991,Pedersen1996}
\begin{equation}
	 \hat{H}_{LK}=\hat{H}_{LK,k}+\hat{H}_{LK,\varepsilon}+V_{off,h}\cdot{\cal I}_{4\times 4} +\kappa {\cal J}_{z}{\cal H}_{z},\\
\label{hlk:complete}
\end{equation}
where $\hat{H}_{LK,k}$ is the kinetic part of the LK Hamiltonian, $\hat{H}_{LK,\varepsilon}$ is the part which
describes the influence of mechanical strain on the valence-band electronic structure, {${\cal I}_{4\times 4}$}
is the {$4\times 4$} identity matrix, {$V_{off,h}$} is the confining potential due to the offset in the valence band,
the diagonal matrix {${\cal J}_{z}={\rm diag}(+3/2,+1/2,-1/2,-3/2)$} contains projections of the angular
momentum {\bf J} (${\it J}= 3/2$) of the zone-center Bloch electrons in the valence band onto the {\it z} -axis,
$\kappa$ is the Luttinger parameter and ${\cal H}_{z}=\hbar\omega_{c}$, where $\omega_{c}=qB/m_{0}$
denotes the cyclotron frequency. Note that effects due to the piezoelectric field are negligible, and are
therefore discarded in our calculations\cite{Ding2010,Fomin2007}. The energies are measured from the
band extrema in the matrix far away from the dot boundary, and the energy axis for the hole states is
directed from the top of the valence band downward to the top of the split-off band.

The kinetic part of the LK model is given by\cite{Chuang1991,Pedersen1996}
\begin{equation}
	\hat{H}_{LK,k}=
	\left\lbrack
	\begin{array}{cccc}
		\hat{P}+\hat{Q} & -\hat{S} & \hat{R} & 0 \\
		-\hat{S}^{\dag} & \hat{P}-\hat{Q} & 0 & \hat{R}\\
		\hat{R}^{\dag} & 0 & \hat{P}-\hat{Q} & \hat{S}\\
		0 & \hat{R}^{\dag} & \hat{S}^{\dag} & \hat{P}+\hat{Q}\\
	\end{array}
	\right\rbrack,\\
\end{equation}
with the matrix elements
\begin{subequations}
	\begin{equation}	 \hat{P}=\frac{\hbar^{2}}{2m_{0}}\left[\frac{1}{2}\left(\hat{k}_{+}\gamma_{1}\hat{k}_{-}+\hat{k}_{-}\gamma_{1}\hat{k}_{+}\right)
		+\hat{k}_{z}\gamma_{1}\hat{k}_{z}\right],
    \end{equation}
    \begin{equation}    
\hat{Q}=\frac{\hbar^{2}}{2m_{0}}\left[\frac{1}{2}\left(\hat{k}_{+}\gamma_{2}\hat{k}_{-}+\hat{k}_{-}\gamma_{2}\hat{k}_{+}\right)
		-2\hat{k}_{z}\gamma_{2}\hat{k}_{z}\right],		
    \end{equation}		
    \begin{equation}     \hat{S}=\frac{\hbar^{2}}{2m_{0}}\sqrt{3}\left(\hat{k}_{-}\gamma_{3}\hat{k}_{z}+\hat{k}_{z}\gamma_{3}\hat{k}_{-}\right),
    \end{equation}
   \begin{equation}         		 \hat{R}=-\frac{\hbar^{2}}{2m_{0}}\sqrt{3}\left({\hat{k}_{-}\overline\gamma}\hat{k}_{-}-\hat{k}_{+}\mu\hat{k}_{+}\right).
    \end{equation}
\label{hk:rmat}
\end{subequations}
Here, $\gamma_{1}$, $\gamma_{2}$, and $\gamma_{3}$ denote the position-dependent Luttinger
parameters, {${\overline\gamma}=(\gamma_{2}+\gamma_{3})/2$}, and
{$\mu=(\gamma_{3}-\gamma_{2})/2$}. For {\it B} oriented along the $z$-axis,
the magnetic vector potential for the hole states in the cylindrical coordinates is expressed by
{${\bf A}_{h}={\bf e}_{\phi}\cdot B\rho_{h} /2$}. Thus, it is straightforward to derive the
following expressions for ${\hat k}_{\pm}$ and ${\hat k}_{z}$:
\begin{subequations}
	\begin{equation}
        \begin{split}
		{\hat k}_{\pm}&=e^{\pm i\phi_{h}}\left[-i\frac{\partial}{\partial\rho_{h}}\pm i\left(-i\frac{1}{\rho_{h}}\frac{\partial}{\partial\phi_{h}}+
		\frac{qB\rho_{h}}{2\hbar}\right)\right]\\
    &=e^{\pm i\phi_{h}}\left({\hat k}_{\rho}\pm i{\hat k}_{\phi}\right),
        \end{split}
    \end{equation}
	\begin{equation}    
 		{\hat k}_{z}=-i\frac{\partial}{\partial z_{h}}.
	\end{equation}
\end{subequations}
The strain-dependent part of the multiband LK Hamiltonian {$\hat H_{LK,\varepsilon}$} is given by
\begin{equation}
	\hat{H}_{LK,\varepsilon}=
	\left\lbrack
	\begin{array}{cccc}
		P_{\varepsilon}+Q_{\varepsilon} & -S_{\varepsilon} & R_{\varepsilon} & 0 \\
		-S_{\varepsilon}^{\dag} & P_{\varepsilon}-Q_{\varepsilon} & 0 & R_{\varepsilon}\\
		R_{\varepsilon}^{\dag} & 0 & P_{\varepsilon}-Q_{\varepsilon} & S_{\varepsilon}\\
		0 & R_{\varepsilon}^{\dag} & S_{\varepsilon}^{\dag} & P_{\varepsilon}+Q_{\varepsilon}.\\
	\end{array}
	\right\rbrack
\end{equation}
Similar to $\hat{H}_{LK,\varepsilon}$, the matrix elements of $\hat{H}_{\varepsilon}$ are written in cylindrical coordinates:
\begin{subequations}
	\begin{eqnarray}
		 P_{\varepsilon}&=&-a_{v}\left(\varepsilon_{\rho\rho}+\varepsilon_{\phi\phi}+\varepsilon_{zz}\right)=-a_{v}\varepsilon_{hyd},\\
		 Q_{\varepsilon}&=&b\left(\varepsilon_{zz}-\frac{\varepsilon_{\rho\rho}+\varepsilon_{\phi\phi}}{2}\right)=b\varepsilon_{b},\\		 
		 S_{\varepsilon}&=&-d\left(\varepsilon_{\rho z}-i\varepsilon_{\phi z}\right)e^{-i\phi_{h}},\\
		 R_{\varepsilon}&=&e^{-2i\phi_{h}}\frac{\sqrt{3}b+d}{4}\left(\varepsilon_{\rho\rho}-\varepsilon_{\phi\phi}-2i\varepsilon_{\rho\phi}\right)\nonumber\\
				 &+&e^{+2i\phi_{h}}\frac{\sqrt{3}b-d}{4}\left(\varepsilon_{\rho\rho}-\varepsilon_{\phi\phi}+2i\varepsilon_{\rho\phi}\right)
\label{he:rmat}.
	\end{eqnarray}
\end{subequations}
Here, {$a_{v}$}, {\it b} and {\it d} are the deformation potentials in the valence band and {$\varepsilon_{b}$} denotes the {\it biaxial strain}.

We briefly discuss the symmetry of the different parts of the LK Hamiltonian in Eq.~(\ref{hlk:complete}). Even though $V_{off,h}$ has axial symmetry, $\hat{H}_{LK,k}$ and $\hat{H}_{LK,\varepsilon}$ lack axial symmetry due to the anisotropy of the bulk electronic structure of the constituent materials and the strain anisotropy, respectively. The difference between $\gamma_2$ and $\gamma_3$ is usually small, and therefore $\mu$ is a small number. If $\mu$ is approximately taken as zero in Eq.~(\ref{hk:rmat}), the $\hat{H}_{LK,k}$ becomes axially symmetric, which is the well-known axial approximation of a multiband ${\bf k}\cdot{\bf p}$ model\cite{Chuang1991}. In such a way, the term responsible for the in-plane anisotropy in the $R$ matrix element of $\hat{H}_{LK,k}$ which is proportional to $e^{+2i\phi_{h}}$ is removed from the multiband Hamiltonian. The term proportional to $e^{+2i\phi_{h}}$ also exists in the $R_\varepsilon$ matrix element of $\hat{H}_{LK,\varepsilon}$, and similar to $\hat{H}_{LK,k}$, is responsible for the lack of axial symmetry in $\hat{H}_{LK,\varepsilon}$. Nevertheless, the difference $\vert\sqrt{3}b-d\vert$ is usually much smaller than $\vert\sqrt{3}b+d\vert$. For example, in ${\rm In_{0.5}Ga_{0.5}As}$ we have $\sqrt{3}b-d=0.9$, whereas $\sqrt{3}b+d=-7.5$\cite{Vurgaftman2001}. If we approximately take $d=\sqrt{3}b$, the term proportional to $e^{+2i\phi_{h}}$ is eliminated, and therefore $\hat{H}_{LK,\varepsilon}$ becomes axially symmetric. Along with the axial approximation of $\hat{H}_{LK,k}$ and the axial symmetry of the confining potential, the approximation $\sqrt{3}b-d \approx 0$ makes the LK model fully axially symmetric.

Therefore, for the axially symmetric RLQD, the application of this model is justified when $\vert\gamma_2-\gamma_3\vert\ll(\gamma_2+\gamma_3)$, $\vert\sqrt{3}b-d\vert\ll\vert\sqrt{3}b+d\vert$. Such a model does not discard mixing  between different hole states due to the off-diagonal matrix elements of $\hat{H}_{LK,\varepsilon}$, and therefore could potentially be more accurate than our previous approach\cite{Tadic2002}, which discarded this influence. The application of the later model was based on the fact that in an axially symmetric quantum dot the off-diagonal terms of the strain tensor are generally small, except close to the dot's lateral boundary\cite{Tadic2002,TadicPRB2002}. Also, $\varepsilon_{\rho\rho}-\varepsilon_{\phi\phi}\approx 0$, except around the top of the rim of the analyzed axially symmetric RLQD, therefore the off-diagonal terms of the strain-dependent Hamiltonian have a small effect on the effective potentials\cite{TadicPRB2002}. This model has the form
\begin{equation}
    \hat{H}_{LK,eff}=\hat{H}_{LK,k}+V_{h,eff}+\kappa {\cal J}_{z}{\cal H}_{z},\\
\label{hlk:approx}
\end{equation}
which is simpler than Eq.\:(\ref{hlk:complete}). Here,
\begin{equation}
    V_{h,eff}={\rm diag}(V_{hh,eff},V_{lh,eff},V_{lh,eff},V_{hh,eff}),
\end{equation}
where
\begin{subequations}
	\begin{equation}
		V_{hh,eff}({\bf r})=-a_v\:\varepsilon_{hyd}({\bf r})-\frac{b}{2}\varepsilon_{hyd}({\bf r})+\frac{3}{2}b\: \varepsilon_{zz}({\bf r})+V_{off,h}({\bf r}),
    \end{equation}
    \begin{equation}
		V_{lh,eff}({\bf r})=-a_v\:\varepsilon_{hyd}({\bf r})+\frac{b}{2}\varepsilon_{hyd}({\bf r})-\frac{3}{2}b\: \varepsilon_{zz}({\bf r})+V_{off,h}({\bf r}).
	\end{equation}
\end{subequations}
The splitting between the HH and LH bands is expressed by the linear combination of the hydrostatic strain {$\varepsilon_{hyd}$} and tensile strain {$\varepsilon_{zz}$}. These expressions for the effective potentials of the heavy- and light-hole states will be subsequently employed to explain how the localization of the hole depends on the RLQD geometry and the associated preferential direction of lattice relaxation.

{$\hat{H}_{LK}$} acts on the multiband envelope function spinor
\begin{equation}
    \begin{split}
	&\varPsi_{n_{h},f_{zh}}^{h}({\bf r}_{h})\\
    &=	 \left[\varPsi_{n_{h},f_{zh},+3/2}^{h},\varPsi_{n_{h},f_{zh},+1/2}^{h},	 \varPsi_{n_{h},f_{zh},-1/2}^{h},\varPsi_{n_{h},f_{zh},-3/2}^{h}\right]^{T},
    \end{split}
\end{equation}
where {$f_{zh}$}={$l_{h}$}+{$j_{zh}$} is the quantum number of the projection
of the total angular momentum {\bf F}={\bf  L}+{\bf J} onto the {\it z} axis,
and {$n_{h}$} is the principal quantum number. Because of the axial symmetry, each
envelope function has the form
\begin{equation}
	\varPsi_{n_{h},f_{zh},j_{z}}^h({\bf r}_{h})=\frac{1}{\sqrt{2\pi}}e^{il_{h}\phi_{h}}\psi_{n_{h},l_{h},j_{z}}(\rho_{h},z_{h}),
\end{equation}

In addition to the multiband LK model given by Eq.\:(11), we employed the
single-band approach which relies on the spherical approximation of the
multiband model. This Hamiltonian has the same form as the Hamiltonian
for the electron states given by Eq.\:(7), with the subscript {\it e} replaced
by {\it h}, and the sign of charge altered ($q\rightarrow-q$). The states of the heavy and light holes are separately computed in the spherical approximation, therefore the respective states of the HH and LH excitons are separately determined. When band mixing is taken into account, the HH and LH excitons mix and produce the {\it multiband exciton}.

\subsection{Exciton states: Multiband approach}

The exciton Hamiltonian has the form
\begin{equation}
	\hat{H}_{x}=\hat{H}_{e}+\hat{H}_{h}+\hat{V}_{C}.
\end{equation}
Here, {$\hat{H}_{e}$} and {$\hat{H}_{h}$} are the single-particle
electron and hole Hamiltonians, given by Eqs.\:(7) and (11), respectively.
{$\hat{V}_{C}=-q^2/\left(4\pi\epsilon\epsilon_{0}|{\bf r}_{e}-{\bf r}_{h}|\right)$}
denotes the Coulomb potential energy of the interacting electron-hole
pair, and the relative permittivity {$\epsilon$} is assumed to correspond to
In(Ga)As, where the electron and hole are mostly localized. Furthermore,
because the Coulomb interaction is spin independent, the electron spin
{$s_{z}$} of the exciton is a good quantum number. As
mentioned above the hole states are classified according to the {\it z} projection
of the hole total angular momentum {$f_{zh}$}, but the
Coulomb interaction mixes different {$f_{zh}$} hole states in the exciton.
However, the motion of the exciton center of mass is axially symmetric,
therefore the {\it z} projection of the exciton total angular momentum
{$f_{zx}=s_{z}+l_{e}-f_{zh}$} is the other good quantum number
of the multiband exciton state. For the given {$s_{z}$} and {$f_{zx}$},
the multiband envelope-function spinor of the exciton state
is expanded in products of the electron and hole envelope functions spinors
\begin{equation}
    \begin{split}
	&\varPsi_{f_{zx},s_{z}}^{x}({\bf r}_{e},{\bf r}_{h})\\
    &=\sum_{n_{e},n_{h}}
	\sum_{l_{e}}c_{l_{e},n_{e},n_{h},s_{z}}
	\varPsi_{n_{e},l_{e}}^{e}\left({\bf r}_{e}\right)
	\varPsi_{n_{h},s_{z}+l_{e}-f_{zx}}^{h}\left({\bf r}_{h}\right).
    \end{split}
\end{equation}
The exciton energy levels are extracted from the secular equation
\begin{equation}
\begin{split}	 
&\left(E_{g}+E_{n_{e},l_{e},s_{z}}+E_{n_{h},s_{z}+l_{e}-f_{zx}}-E_{x}\right)\\	 &\times\cdot\delta_{l_{e},l'_{e}}\cdot\delta_{n_{e},n'_{e}}\cdot\delta_{n_{h},n'_{h}}\\
	&+\sum_{n_{e},n_{h}}\sum_{l_{e}}
	\left<\varPsi_{n'_{e},l'_{e}}^{e}({\bf r}_{e})\varPsi_{n'_{h},s_{z}+l'_{e}-f_{zx}}^{h}({\bf r}_{h})
	\left|V_{C}\right|\varPsi_{n_{e},l_{e}}^{e}({\bf r}_{e})\right.\\
&\times\left.\varPsi_{n_{h},s_{z}+l_{e}-f_{zx}}^{h}({\bf r}_{h})\right>=0.
\end{split}
\end{equation}
Here, {$E_{g}$} denotes the band gap energy of the matrix semiconductor.
We note that the secular equations of the {$\left|f_{zx},n_{x},\uparrow\right>$} and
{$\left|f_{zx}-1,n_{x},\downarrow\right>$} exciton states are equal,
while their energies differ by {$g_{e}^*\mu_{B}B$}. The energy of
the lowest optically active (bright) exciton state is conveniently denoted by
$E_{x,1}$. For zero magnetic field, due to electron spin degeneracy and
Kramers degeneracy of the hole states, the exciton states are arranged in quadruplets.

\subsection{Exciton states: Single-band approximation}

Because of the larger effective mass of the heavy hole, the exciton
ground state is the HH exciton. The single-band approximation does not
take into account the effects of spin-orbit coupling on the valence-band
states, and therefore the spins of both the electron and the hole are
good quantum numbers for the HH exciton. Also, the {\it z} component
of the total orbital momentum {$l_{x}=l_{e}+l_{h}$} is a good quantum number
for the exciton. For the given {$l_{x}$} and either {$s_{z}$} or {$j_{z}$},
the exciton envelope function is expanded in products of the electron
and hole envelope functions
\begin{equation}
	\varPsi_{l_{x}}^{x}({\bf r}_{e},{\bf r}_{h})=\sum_{n_{e},n_{h}}
	\sum_{l_{e}}c_{n_{e},n_{h},l_{e}}
	\varPsi_{n_{e},l_{e}}^{e}\left({\bf r}_{e}\right)
	\varPsi_{n_{h},l_{x}-l_{e}}^{h}\left({\bf r}_{h}\right).
\end{equation}
It is straightforward to show that the spin-dependent Zeeman terms
contribute to the exciton energy by {$E_{s_{z},j_{z}}^{\rm spin}=g_{s_{z},j_{z}}^*\mu_{B}B/2$},
where {$g_{s_{z},j_{z}}^*=-{\rm sign}(s_{z}-j_{z})$}
{$\times(g_{h}^*-(-1)^{|s_{z}-j_{z}|}g_{e}^*)$}
is the effective exciton {\it g} factor. Hence, {$E_{s_{z},j_{z}}^{\rm spin}$}
could be subtracted from the exciton energy, which simplifies the calculation.

The expansion of Eq.\:(25) leads to the secular equation for the HH
exciton energy
\begin{equation}
    \begin{split}
	&(E_{g}+E_{n_{e},l_{e}}+E_{n_{h},l_{x}-l_{e}}-E_{x}\:)\\	 &\times\cdot\delta_{l_{e},l'_{e}}\cdot\delta_{n_{e},n'_{e}}\cdot\delta_{n_{h},n'_{h}}\\
	&+\sum_{n_{e},n_{h}}\sum_{l_{e}}
	\left<\varPsi_{n'_{e},l'_{e}}^{e}({\bf r}_{e})\varPsi_{n'_{h},l_{x}-l'_{e}}^{h}({\bf r}_{h})
	\left|V_{C}\right|\varPsi_{n_{e},l_{e}}^{e}({\bf r}_{e})\right.\\
&\times\left.\varPsi_{n_{h},l_{x}-l_{e}}^{h}({\bf r}_{h})\right>=0.
    \end{split}
\end{equation}
For zero magnetic field, because of double spin degeneracy of the
electron and hole, and the orbital degeneracy of the single-particle states,
the exciton states are arranged in octuplets, except the $l_{x}=0$ states which are fourfold degenerate.

\subsection{Exciton radius}

We define the {\it average in-plane exciton radius} by
\begin{equation}
\begin{split}	 &\rho_{\parallel}^2=\int_{\Omega_{e}}\int_{\Omega_{h}}{\varPsi^{x}}^\dag({\bf r}_{e},{\bf r}_{h})\cdot	 \left(\rho_{e}^2+\rho_{h}^2-2\rho_{e}\rho_{h}\cos(\phi_{e}-\phi_{h})\right)\\
&\times{\varPsi^{x}}({\bf r}_{e},{\bf r}_{h})d{\bf r}_{h}d{\bf r}_{e}.
\end{split}
\end{equation}
Here, {${\varPsi^{x}}({\bf r}_{e},{\bf r}_{h})$} denotes either the exciton
envelope function spinor in the multiband model or the exciton envelope
function in the single-band approximation. For the case of the multiband
exciton, inserting the expansion of Eq.\:(23) into Eq.\:(27) leads to
\begin{equation}
\begin{split} &\rho_{\parallel}^2=\sum_{l_{e}}\sum_{n_{e},n'_{e}}\sum_{n_{h},n'_{h}}
	[c_{l_{e},n_{e},n_{h},s_{z}}c_{l_{e},n'_{e},n'_{h},s_{z}}\\ &\times\left\langle l_{e},n'_{e}\left\vert\rho_{e}^2\right\vert l_{e},n_{e}\right\rangle\cdot\delta_{n_{h},n'_{h}}\\
	&+c_{l_{e},n_{e},n_{h},s_{z}}c_{l_{e},n'_{e},n'_{h},s_{z}}\\	 &\times\left<s_{z}+l_{e}-f_{zx},n'_{h}\left|\rho_{h}^2\right|s_{z}+l_{e}-f_{zx},n_{h}\right>\cdot\delta_{n_{e},n'_{e}}\\
&-c_{l_{e},n_{e},n_{h},s_{z}}c_{l_{e}-1,n'_{e},n'_{h},s_{z}}
	\left<l_{e}-1,n'_{e}\left|\rho_{e}\right|l_{e},n_{e}\right>\\	 &\times\left<s_{z}+l_{e}-f_{zx}-1,n'_{h}\left|\rho_{h}\right|s_{z}+l_{e}-f_{zx},n_{h}\right>\\
	&-c_{l_{e},n_{e},n_{h},s_{z}}c_{l_{e}+1,n'_{e},n'_{h},s_{z}} \left<l_{e}+1,n'_{e}\left|\rho_{e}\right|l_{e},n_{e}\right>\\	 &\times\left<s_{z}+l_{e}-f_{zx}+1,n'_{h}\left|\rho_{h}\right|s_{z}+l_{e}-f_{zx},n_{h}\right>].
\end{split}
\end{equation}
The expression for {$\rho_{\parallel}$} of the HH exciton state is similarly derived,
\begin{equation}
\begin{split}
&\rho_{\parallel}^2=\sum_{l_{e}}\sum_{n_{e},n'_{e}}\sum_{n_{h},n'_{h}}
	[c_{l_{e},n_{e},n_{h}}c_{l_{e},n'_{e},n'_{h}}\\	 &\times\left<l_{e},n'_{e}\left|\rho_{e}^2\right|l_{e},n_{e}\right>\cdot\delta_{n_{h},n'_{h}}\\
&+c_{l_{e},n_{e},n_{h}}c_{l_{e},n'_{e},n'_{h}}\\	 &\times\left<l_{x}-l_{e},n'_{h}\left|\rho_{h}^2\right|l_{x}-l_{e},n_{h}\right>\cdot\delta_{n_{e},n'_{e}}\\
&-c_{l_{e},n_{e},n_{h}}c_{l_{e}-1,n'_{e},n'_{h}}	 \left<l_{e}-1,n'_{e}\left|\rho_{e}\right|l_{e},n_{e}\right>\\	 &\times\left<l_{x}-l_{e}+1,n'_{h}\left|\rho_{h}\right|l_{x}-l_{e},n_{h}\right>\\
&-c_{l_{e},n_{e},n_{h}}c_{l_{e}+1,n'_{e},n'_{h}}	 \left<l_{e}+1,n'_{e}\left|\rho_{e}\right|l_{e},n_{e}\right>\\	 &\times\left<l_{x}-l_{e}-1,n'_{h}\left|\rho_{h}\right|l_{x}-l_{e},n_{h}\right>].
\end{split}
\end{equation}
We previously found that {$\rho_{\parallel}$} exhibits oscillations with
magnetic field whose period is similar to oscillations in the exciton energy\cite{Arsoski2012,Tadic2011}.

We also define the {\it average single-particle radius} and {\it vertical position}
for the given exciton state,
\begin{equation}
	 \left<\rho_{e(h)}\right>=\int_{\Omega_{e}}\int_{\Omega_{h}}{\varPsi^{x}}^\dag({\bf r}_{e},{\bf r}_{h})\cdot
	\rho_{e(h)}\cdot{\varPsi^{x}}({\bf r}_{e},{\bf r}_{h})d{\bf r}_{h}d{\bf r}_{e}.
\end{equation}
and
\begin{equation}
	 \left<z_{e(h)}\right>=\int_{\Omega_{e}}\int_{\Omega_{h}}{\varPsi^{x}}^\dag({\bf r}_{e},{\bf r}_{h})\cdot
	z_{e(h)}\cdot{\varPsi^{x}}({\bf r}_{e},{\bf r}_{h})d{\bf r}_{h}d{\bf r}_{e},
\end{equation}
respectively. For the concentric 1D ring with the radii
{$<\rho_{e}>$} and {$<\rho_{h}>$}, the exciton polarization
depends on {$<\rho_{e}>-<\rho_{h}>$}\cite{Govorov2002}. Furthermore,
these two radii determine the effective exciton orbital radius {$\lambda_{x}$},
which is for the HH exciton defined by\cite{Ding2010}
\begin{equation}
	 \lambda_{x}=\left[\frac{m_{e}+m_{h}}{m_{e}/<\rho_{h}>^2+m_{h}/<\rho_{e}>^2}\right]^\frac{1}{2}.
\end{equation}
Together with {$<\rho_{e}>-<\rho_{h}>$} it determines the {\it effective surface area}
(as the one between concentric 1D rings) threaded by the magnetic flux. Because the hole
effective mass is larger, {$\lambda_{x}$} mostly reflects the behavior of {$<\rho_{e}>$} with
magnetic field, as demonstrated in Refs.~\onlinecite{Ding2010} and \onlinecite{Li2011} for the case
where compositional intermixing between the ring and the matrix is present.
Furthermore, by using a simple analytical argument\cite{Barticevic2006}, it could
be shown that the diamagnetic shift coefficient\cite{Ding2010} is dominated by
a term proportional to {$\lambda_{x}^2/\mu_{eh}$}, where
{$\mu_{eh}=m_{e}m_{h}/({m_{e}+m_{h}})$}. The dependence of {$\lambda_{x}$}
on {\it B} describes part of the variation of the exciton energy
around its average parabolic dependence on magnetic field.

\subsection{Estimation of the oscillations of the exciton energy }

To resolve the AB oscillations in the exciton energy, the authors of Ref.~\onlinecite{Grochol2006}
proposed to plot the second derivative
{$d^2E_{x,1}/dB^2$} as a function of {\it B}. We propose here an
alternative way to estimate the period and magnitude of the exciton
energy oscillation. For that purpose, we fit the energy of the lowest  optically active state $E_{x,1}$
as a function of {\it B} by a polynomial of the fourth order,
\begin{equation}
	\left<E_{x,1}\right>=\sum_{i=0}^{4} c_{i}B^i,
\end{equation}
where {$c_{i}$}'s are the best-fit parameters. We found
that the fourth order polynomial describes reasonably well the
correction from a purely parabolic dependence that is found for the 1D
model. The exciton energy residual,
\begin{equation}
	\delta E_{x,1}=E_{x,1}-\left<E_{x,1}\right>,
\end{equation}
is expected to exhibit the AB oscillations.

An additional figure of merit of the exciton state is the
{\it in-plane Coulomb potential energy}, which is defined by
\begin{equation}
	 V_{C\parallel}=-\frac{q^2}{4\pi\epsilon\epsilon_{0}\rho_{\parallel}}.
\end{equation}
It will be demonstrated that for the lowest-energy bright exciton state {$V_{C\parallel}$} exhibits oscillations
around a linear function
\begin{equation}
	\left<V_{C\parallel}\right>=d_{0}+d_{1}B,
\end{equation}
where {$d_{i}$}'s are the best-fit parameters. The oscillatory behavior
of {$V_{C\parallel}$} is then resolved by computing the residual
\begin{equation}
	\delta V_{C\parallel}=V_{C\parallel}-\left<V_{C\parallel}\right>.
\end{equation}

\subsection{Oscillator strength and photoluminescence intensity}

For the fully opened quantum rings, we previously found that
oscillations of the exciton energy levels occur together with
oscillations in the oscillator strength for exciton recombination\cite{Tadic2011}.
For the {\it i}-th exciton state, this oscillator strength is given by
\begin{equation}
	f_{x,i}=\frac{2}{m_{0}E_{x,i}}\big| \left< u_{c0} \left| {\boldsymbol \varepsilon}\cdot {\bf p} \right| u_{v0} \right> \big|^2 \left| M \right|^2.
\end{equation}
Here, {$\boldsymbol \varepsilon$} denotes the unit vector of light polarization,
{$u_{c0}$} and {$u_{v0}$} are the periodic parts of the Bloch
functions of the electron in the conduction and valence band,
respectively, and {\it M} denotes the transition matrix element
between the envelope functions \cite{Efros1996}
\begin{equation}
	M=\int_{\Omega_{e}}\int_{\Omega_{h}}\delta({\bf r}_{e}-{\bf r}_{h})\cdot{\varPsi^{x}}({\bf r}_{e},{\bf r}_{h})d{\bf r}_{e}d{\bf r}_{h}.
\end{equation}
We assume that {$\boldsymbol \varepsilon$} is oriented in the {\it xy}
plane (in-plane polarized light), thus the matrix element squared
between the zone center Bloch states is given by
\begin{equation}
	\big| \left< u_{c0} \left| {\boldsymbol \varepsilon}\cdot {\bf p} \right| u_{v0} \right> \big|^2=
	\frac{m_{0}^2P^2}{2\hbar^2}\cdot\delta_{|s_{z}-j_{z}|,1},
\end{equation}
where {\it P} denotes the Kane interband matrix element.

The exciton recombination is strongly polarization sensitive. By
inspecting the spin part of the Bloch functions for the in-plane light
polarization, one may find that the valence band functions
{$\varPsi_{n_{h},f_{zx},+3/2}$} and {$\varPsi_{n_{h},f_{zx},-1/2}$}
are the spin-up states and therefore can optically couple with conduction-band spin-up
states, whereas {$\varPsi_{n_{h},f_{zx},+1/2}$} and {$\varPsi_{n_{h},f_{zx},-3/2}$}
can optically couple with spin-down states in the conduction band. Therefore, the spin
selection rule for the exciton recombination is {$|s_{z}-j_{z}|=1$}.
Furthermore, only equal orbital momenta of the electron and the hole
give a non-zero contribution to the transition matrix element in Eq.\:(39).
Therefore, {$|f_{zx}|=1$} are the only bright exciton states for the
in-plane light polarization. Similarly, in the single-band model only
transitions between states of the same spin and {$l_{x }$}=0 are
optically active.

Even though our main interest are oscillations of the exciton
ground-state energy with magnetic field, at finite temperature {\it T}
there exists a finite probability of population of higher exciton
states, which could smear out the oscillations observed in the ground
state. A measurable quantity at finite temperature is the
photoluminescence intensity\cite{Degani2008}, which we define to be
dimensionless, and represent it by the oscillator strength thermally averaged
over all exciton states,
\begin{equation}
	I_{PL}=\frac{\sum_{i}f_{x,i}\exp(-E_{x,i}/k_{B}T)}
	{\sum_{i}\exp(-E_{x,i}/k_{B}T)}.
\end{equation}

\section{Numerical results and discussion}

We assumed that the RLQD, whose shape is shown in Fig.\:1(a), is made of
In$_{0.5}$Ga$_{0.5}$As which is surrounded by the GaAs matrix
\cite{Ding2010,Teodoro2010,Fomin2007,Granados2003}.
The values of the parameters of the function which
describe the RLQD shape are extracted from the measurements in Ref.~\onlinecite{Fomin2007}, and are given in Sec. II.B. For the strain computation, the lateral boundary of the wetting layer is assumed to be circular with radius 100 nm. The calculations are performed for {$\eta$}=0.8, 1.0, and 1.2,
which are the cases depicted in Fig.\:1(b). As noted in Sec. I, in addition to these RLQD's, we compute the exciton states in a quantum ring of cross section shown in Fig.\:1(b), which is described by Eq.~(\ref{ring:open}). This ring is assumed to have a height {$h=h_{M}$} =3.6 nm, whereas the inner and outer radii at half the ring height amount to {$R_{1}$}=8 nm and {$R_{2}$}=15 nm, respectively,  thus $\sigma=4.95$ nm.

The values of the lattice constants, the deformation potentials,
the Luttinger parameters, the energy gaps in InAs and GaAs, and the
bowing parameters for In(Ga)As are all extracted from Ref.~\onlinecite{Vurgaftman2001}. The HH and LH masses determined from these values are in good agreement with the results from measurements\cite{Tanaka1999}. Therefore, we employed the following set of interpolating formulas to compute the values of the Luttinger parameters in the In(Ga)As alloy

\begin{widetext}
\begin{subequations}
	\begin{eqnarray}
	\gamma_{1}^{\rm In(Ga)As}&=&\frac{1}{2}\left(\frac{1-x}{\gamma_{1}^{\rm GaAs}-2\gamma_{2}^{\rm GaAs}}+\frac{x}{\gamma_{1}^{\rm InAs}-2\gamma_{2}^{\rm InAs}}
	+0.145x(1-x)\right)^{-1}\nonumber\\
	&+&\frac{1}{2}\left(\frac{1-x}{\gamma_{1}^{\rm GaAs}+2\gamma_{2}^{\rm GaAs}}+\frac{x}{\gamma_{1}^{\rm InAs}+2\gamma_{2}^{\rm InAs}}-0.0202x(1-x)\right)^{-1},\\
	\gamma_{2}^{\rm In(Ga)As}&=&-\frac{1}{4}\left(\frac{1-x}{\gamma_{1}^{\rm GaAs}-2\gamma_{2}^{\rm GaAs}}+\frac{x}{\gamma_{1}^{\rm InAs}-2\gamma_{2}^{\rm InAs}}
	+0.145x(1-x)\right)^{-1}\nonumber\\
	&+&\frac{1}{4}\left(\frac{1-x}{\gamma_{1}^{\rm GaAs}+2\gamma_{2}^{\rm GaAs}}+\frac{x}{\gamma_{1}^{\rm InAs}+2\gamma_{2}^{\rm InAs}}-0.0202x(1-x)\right)^{-1},\\
	\gamma_{3}^{\rm In(Ga)As}&=&\gamma_{2}^{\rm In(Ga)As}+(\gamma_{3}^{\rm GaAs}-\gamma_{2}^{\rm GaAs})(1-x)\nonumber\\
	&+&(\gamma_{3}^{\rm InAs}-\gamma_{2}^{\rm InAs})x-0.481x(1-x).
	\end{eqnarray}
\end{subequations}
\end{widetext}

Such determined Luttinger parameters are consistent with the results for the {$\rm Ga_{0.47}In_{0.53}As$} alloy\cite{Alavi1980}. Furthermore, we note that $\vert\sqrt{3}b-d\vert=0.9$, is much smaller than $\vert\sqrt{3}b+d\vert=7.5$,  and $\vert\gamma_{2}-\gamma_{3}\vert\ll(\gamma_{2}+\gamma_{3})$, thus the use of the approximate forms of $\hat{H}_{LK,k}$ and $\hat{H}_{LK,\varepsilon}$ is justified for computing the hole states by the multiband model.

In our calculations, the effective Land\'e {\it g} factor was taken to be equal  $g_{e}^*=-0.44$, the Luttinger parameter {$\kappa= 1.72$}, which are the values that correspond to the GaAs matrix\cite{Lawaetz1971}, and were found to lead to a better agreement with experimental measurements\cite{Bayer2004}. The heavy-hole states were computed by using the value of the effective masses as in Ref.~\onlinecite{Ding2010}, whereas the heavy-hole effective {\it g} factor $g_{h}^*=-3.56$ was selected such that the effective exciton {\it g} factor $g_{ex}=-4$, as derived from the recent photoluminescence measurements on  In(Ga)As/GaAs and InAs/InP quantum dots\cite{Nakaoka2004,Kleemans2009}. The negative exciton $g$ factor agrees with the theory of the multiband exciton states\cite{Nakaoka2004,Pryor2006}, and implies that the lowest-energy hole state in the single-band approximation is a spin-up HH state.

The conduction-band offset in the In(Ga)As/GaAs system is assumed to amount to 82\% of the band-gap difference between In(Ga)As and GaAs.\cite{Arsoski2012} The value of the relative permittivity is determined from {$\epsilon (x)=15.1-2.87x+0.67x^2$}(Ref.\:\onlinecite{Ding2010}), for the assumed molar fraction value {\it x}=0.5. The exciton states are computed in the range of {\it B} from 0 to 30 T. The ambient temperature {\it T}= 4.2 K is assumed.

In addition to the effective potentials, the electron and hole localization could be related to the band edges, which are defined  by\cite{Tadic2002}
\begin{subequations}
	\begin{equation}
        E_{cb}({\bf r})=V_e({\bf r})=a_c\varepsilon_{hyd}({\bf r})+V_{off,e}({\bf r}),
\label{vcb:loc}
    \end{equation}
    \begin{equation}
        \begin{split}
		&E_{hh}({\bf r})=V_{off,h}({\bf r})+P_{\varepsilon}+{\rm sign}(Q_{\varepsilon})\\
&\times\sqrt{\mid Q_{\varepsilon}\mid^{2}+\mid S_{\varepsilon}\mid^{2}+\mid R_{\varepsilon}\mid^{2}},
\label{vhh:loc}
    \end{split}
    \end{equation}
    \begin{equation}
        \begin{split}		
        &E_{lh}({\bf r})=V_{off,h}({\bf r})+P_{\varepsilon}-{\rm sign}(Q_{\varepsilon})\\
        &\times\sqrt{\mid Q_{\varepsilon}\mid^{2}+\mid S_{\varepsilon}\mid^{2}+\mid R_{\varepsilon}\mid^{2}}
\label{vlh:loc},
    \end{split}
    \end{equation}
\end{subequations}
for the conduction-band electron, the heavy hole, and the light hole, respectively.

\begin{figure*}
        \begin{center}
	\includegraphics[width=17.8 cm]{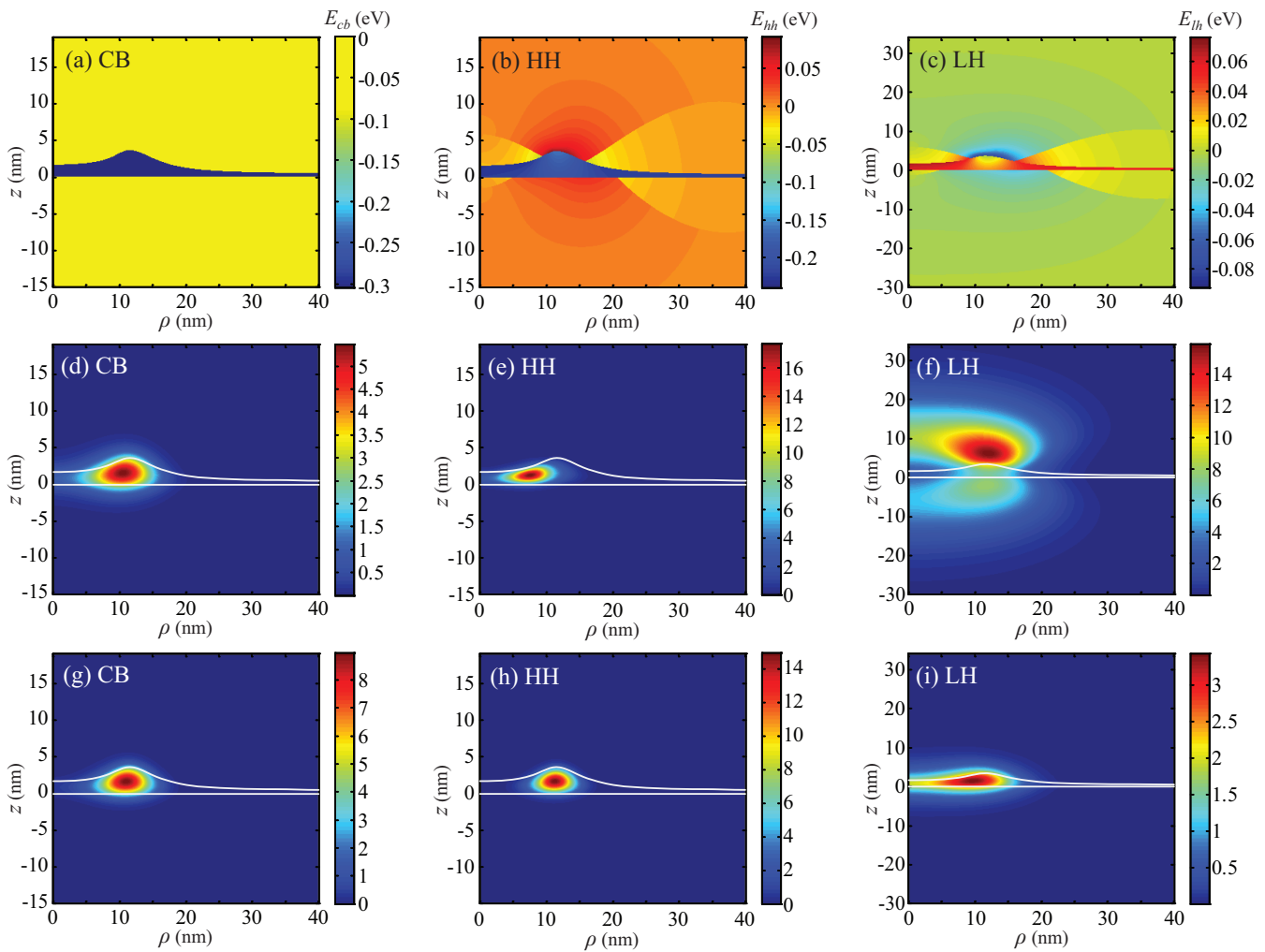}
	\caption{\label{fig2}(Color online) (Upper panel) Contour plots of the band edges in the In(Ga)As/GaAs RLQD for {$\eta=1$}: (a) {$E_{cb}$}, (b) {$E_{hh}$}, and (c) {$E_{lh}$}. (Middle panel) The probability densities of the ground single-particle states which correspond to the band edges in the upper panel for $B=0$: (d) the electron, (e) the heavy hole, and (f) the light hole. (Lower panel) The probability densities of the ground single-particle states for $B=0$, but with strain discarded: (g) the electron, (h) the heavy hole, and (i) the light hole.}
    \end{center}
\end{figure*}

The band edges in the conduction, HH, and LH bands for {$\eta=1$} are shown in the upper panel of Fig.~2[(Figs. 2(a-c)].  These diagrams are generated by using the results from calculations for the strain distribution and Eqs. (43a)-(43c). Hydrostatic strain is constant inside the dot and zero in the matrix, therefore the
conduction band edge, which is shown in Fig.\:2(a), is piecewise constant. Close to the rim top the dot lateral dimension is much smaller than its vertical dimension. Thus, the lattice is mainly relaxed in the lateral plane, whereas closer to the bottom of the RLQD there is no preferential direction for the lattice relaxation. Also, the quantum well for the heavy hole
is deeper close to the dot center than in the rim. As a consequence, the heavy hole is confined in a shallower effective potential well at the top of the rim than at the bottom [see Fig.\:2(b)]. Furthermore, due to strain, the barrier for the LH states is erected inside the dot, and the shallow effective potential well for the LH states is formed outside the dot, which is shown in Fig.~2(c). One might infer that because of confinement of the HH and LH states in the different regions of the strained structure, band mixing is reduced, which is advantageous for the application of the single-band approximation when computing the exciton states.\cite{Ding2010}

Along with the variations in the band edge, the probability densities of the ground electron, HH, and LH states for the cases $\eta=1$ and $B=0$ are displayed in Fig.~2. The middle panel [Figs.~2(d-f)] displays the electron and hole localization in the strained RLQD, whereas the lower panel [Figs.~2(g-i)] show the states in the absence of strain.
The localization of the electron in the ground state does not qualitatively depend on the presence of strain, which could be inferred from a comparison of  Figs.~2(d) and 2(g). However, strain leads to a deeper effective potential well for the heavy hole in the crater than in the rim, therefore the HH ground state is confined closer to the RLQD center [compare Figs.~2(d) and 2(e)], i.e. there is an obvious tendency of the heavy-hole probability density in the strained RLQD to leak towards the center. On the other hand, the heavy-hole in the unstrained structure is mainly localized in the rim, like the electron [compare Figs.~2 (g) and 2(h)]. The peculiar effective potential that confines  the light hole shown in Fig.~2(c) favors the LH localization above and below the dot, as illustrated in Fig.\:2(f). However, when strain is not taken into account, the LH state becomes mainly localized in the dot, as shown in Fig.~2(i). Different localization of the LH states implies that band mixing has considerably different effects on the hole states in strained and unstrained RLQD's.

To demonstrate how the effective potential for the heavy hole varies with the height of the RLQD rim, we plot in Fig.\:3(a) {$E_{hh}$} as a function of {$\rho$} at height {\it z}=0.8 nm, which is half the height of the RLQD crater. It is obvious that when the rim height decreases, the potential well in the inner layer becomes shallower, and the barrier in the rim becomes lower. The probability densities of the ground $f_z=+3/2$ states computed by the LK model in both the strained and unstrained RLQD's, which are displayed in Figs.~3(b) and 3(c), are similar to the probability densities of the respective HH states, shown in Figs.~2(e) and 2(h), respectively. However, due to the peculiar in-plane variation of $E_{hh}$ shown in Fig.~3(a), the probability density of the multiband hole state in the {\it strained} structure leaks more effectively to the RLQD center. Therefore, when the rim height increases the hole localization inside the rim is reduced when strain is present. On the other hand, because the  electron effective potential is stepwise, the influence of increasing the rim height on the electron localization inside the rim is more effective. Therefore, such a difference is advantageous for increasing the exciton polarization, and in turn it could increase the excitonic AB oscillations.

\begin{figure}
        \begin{center}
	\includegraphics[width=7 cm]{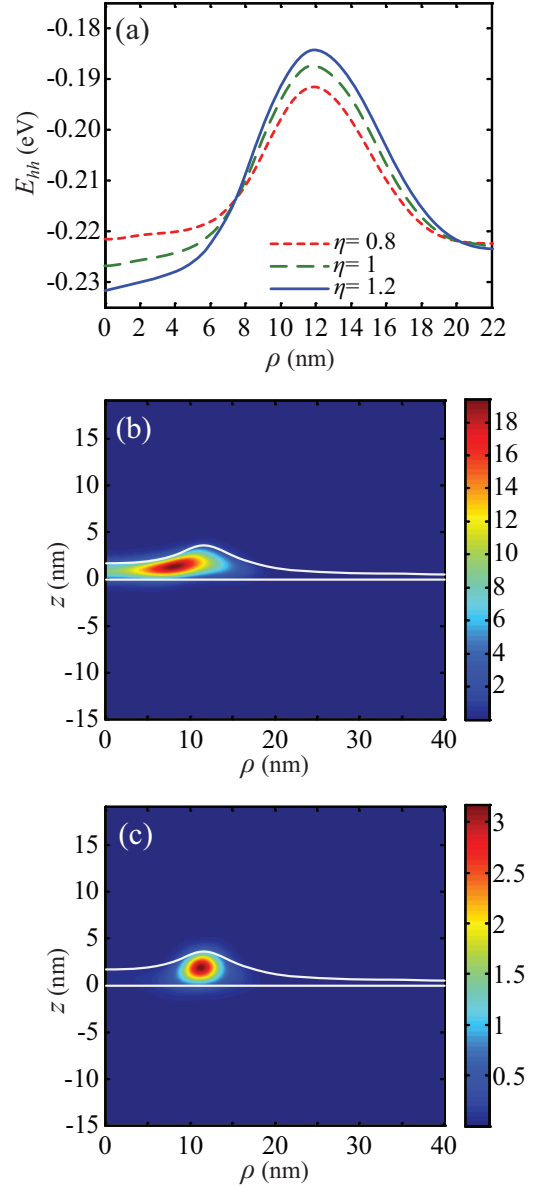}
	\caption{\label{fig3}(Color online) (a) {$E_{hh}$} variation with {$\rho$} for {$z=0.8$} nm. (b) The probability density of the hole ground state when  band mixing is taken into account. (c)  The probability density of the hole ground state when band mixing is taken into account, but strain is discarded.}
    \end{center}
\end{figure}

\begin{figure}
        \begin{center}
	\includegraphics[width=7 cm]{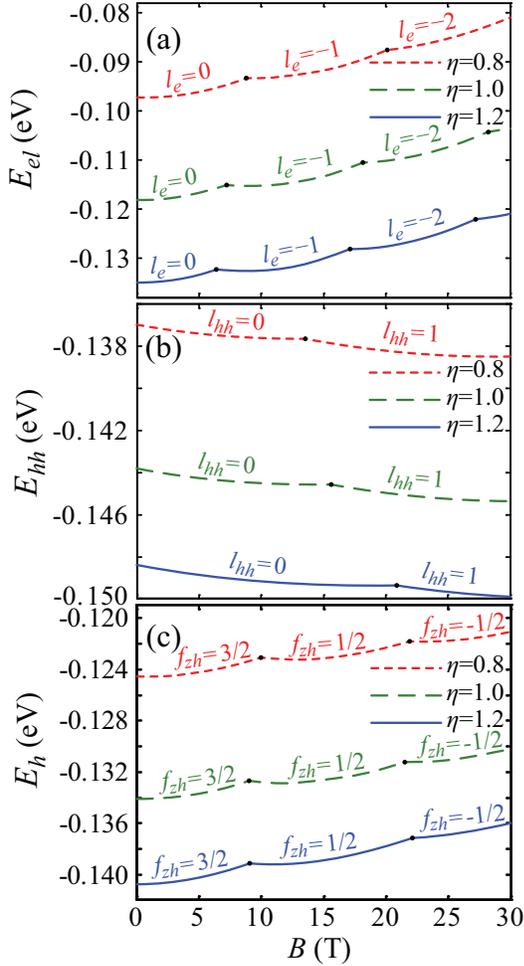}
	\caption{\label{fig4}(Color online) The ground state energy of: (a) the electron, (b)
	the heavy hole within the single-band approximation, and (c) the hole
	with band mixing taken into account in the LK model. The dotted {\color{mred} {red}}
	lines denote the case {$\eta=0.8$}, the dashed {\color{mgreen} green} lines are the
	states in the RLQD for {$\eta=1.0$}, and the solid {\color{mblue}blue} lines
	depict the case {$\eta=1.2$}. The angular momentum transitions are
	labeled by dots, and the orbital (angular) quantum numbers of the
	electron and hole ground states are explicitly indicated.}
    \end{center}
\end{figure}

The single-particle energy levels as function of the magnetic field for
{$\eta=0.8$} (dotted {\color{mred}red} lines), {$\eta=1.0$} (dashed {\color{mgreen}green} lines),
and {$\eta=1.2$} (solid {\color{mblue}blue} lines), are shown in
Fig.\:4. For {$\eta=1.2$} the rim of the RLQD is high enough to
establish that the electron is localized mostly in the rim. For this
case, the orbital momentum transitions are at {$B_{1}$}=6.44 T,
{$B_{2}$}=17.2 T, and {$B_{3}$}=27.3 T. One might note that
the orbital momentum transitions in 1D rings occur when an odd
multiple of half flux quantum threads the ring\cite{Lee2004}.
Therefore, they are arranged according to {$B_{i}/B_{1}=(2i-1)$},
{\it i}=2,3,\ldots, where {$B_{1}$} is the magnetic field of the first orbital
momentum transition. This relation is approximate for {$\eta=1.2$},
whereas for smaller {$\eta$} deviation from the 1D ring case
becomes larger. It is ascribed to a larger localization in the inner
layer when the rim's volume decreases. Consequently, the orbital
momentum transitions shift towards higher {\it B} values.

Because the effective potential well for the HH states is deeper inside
the inner layer than in the rim [see Fig.\:3(a)], the hole is localized
closer to the RLQD center than the electron. Therefore, the first
orbital momentum transition for {$\eta=1.2$} is at much larger
\textit{B} than for the electron [compare Figs.\:4(a) and 4(b)]. For
RLQDs with a smaller rim, which are the cases {$\eta=0.8$} and
{$\eta=1.0$}, the hole is confined in a smaller volume, but the
effective potential barrier for the heavy hole in the rim lowers due to strain. A consequence of the later effect is an increase of the average heavy-hole radius. Therefore, the first orbital momentum transition between the heavy-hole states shifts towards smaller {\it B} values when $\eta$ decreases.

One might note that the ground HH energy level in Fig.\:4(b) lowers below the $B=0$ value, which is a consequence of a too large Zeeman splitting in the single-band approximation of the hole states. However, mixing between the hole bands due to the off-diagonal kinetic terms of the LK model opposes effects of the Zeeman splitting. Furthermore, the off-diagonal strain-dependent terms affects the angular momentum transition, as Fig.\:4(c) shows. As a matter of fact, the diagrams of the electron and hole states in Figs.\:4(a) and 4(c) appear qualitatively similar.  On the other hand, a comparison between Figs.\:4(b) and 4(c) demonstrates that the single-band and multiband models of the hole states exhibit considerable qualitative discrepancy, i.e., band mixing has profound effects on the Aharonov-Bohm oscillations of the hole ground energy level. Therefore, the changes of the hole localization due to the peculiar effective potential variation with the rim height shown in Fig.\:3(a) cannot straightforwardly explain how the single-particle energy levels in the LK model are affected by the increase of the RLQD height.

\begin{figure*}
        \begin{center}
	\includegraphics[width=14cm]{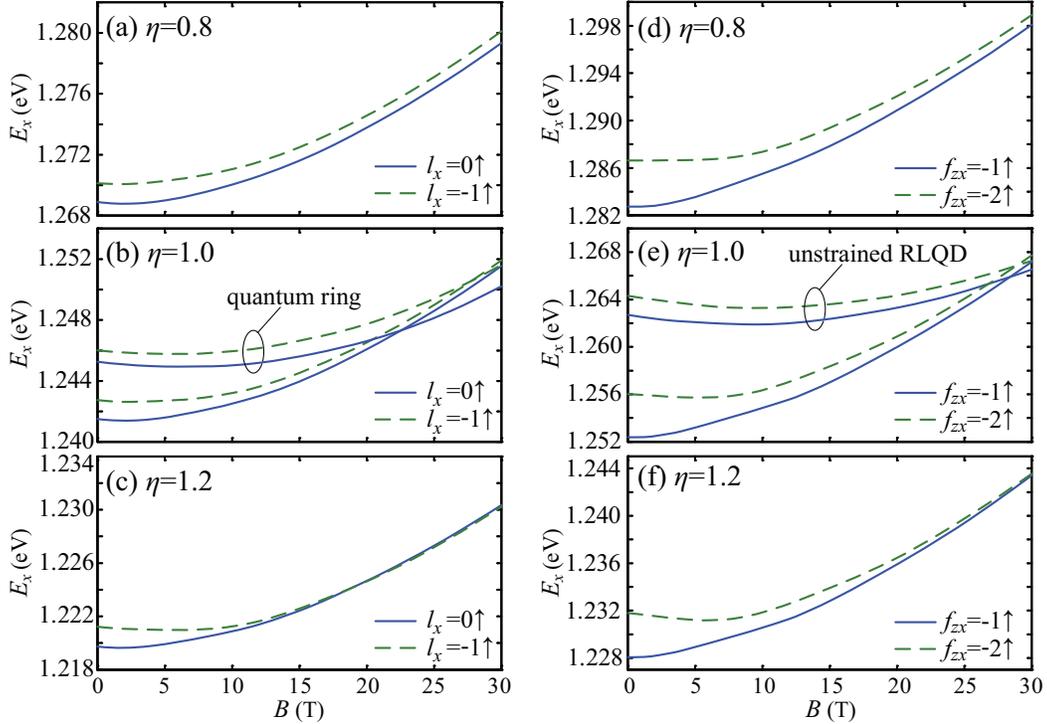}
	\caption{\label{fig5}(Color online) Variations of the lowest-energy levels of a few
	orbital (angular) momenta for the electron spin up with magnetic field.
	(Left panel.) The HH exciton energy levels for: (a) {$\eta=0.8$}
	(b) {$\eta=1.0$} (the exciton energy levels decreased
	by 45 meV in the fully opened quantum ring are also shown), and
	(c) {$\eta=1.2$}. (Right panel.) The multiband exciton energy levels
	for: (d) {$\eta=0.8$}, (e) {$\eta=1.0$} (the exciton energy levels increased by 125 meV in the same RLQD but with strain discarded are also shown), and (f) {$\eta=1.2$}.}
      \end{center}
\end{figure*}

Figure\:5 shows how the low exciton energy levels vary with magnetic field.
The HH exciton energies determined by the single-band approximation are
shown in the left panel, whereas the results of the multiband
calculation are displayed in the right panel. For both models, the presence of small oscillations of the exciton ground-state energy around a parabolic function are found in Fig.~5 when {$\eta$} increases. This is similar to experiments\cite{Ding2010,Teodoro2010} and our previous work on fully opened quantum rings.\cite{Arsoski2012,Tadic2011} For $\eta=0.8$ and $\eta=1.0$, the exciton ground energy level is optically active (either $l_x=0$ or $f_{zx}=-1$) in the whole explored range of $B$ from 0 to 30 T. On the other hand, for $\eta=1.2$ the $l_x=0$ and $l_x=-1$ levels cross each other at 19.1 T. This crossing is a consequence of the decreased Coulomb interaction between the electron and hole due to increased exciton polarization when the rim height increases, which was previously demonstrated in Figs.~2 and 3. However, the  multiband exciton does not exhibit the angular momentum transitions in the ground state shown in Fig.~5(f). The difference between the HH and multiband exciton ground states shown in Figs.~5(c) and 5(f) is related to both band mixing and locations of the single-particle orbital (angular) momentum transitions. In the single-band approach, the electron orbital momentum transitions are misplaced with respect to the hole orbital momentum transitions. Therefore, a change of the exciton orbital momentum $l_x=l_e+l_h$ is possible when $B$ varies. When band mixing is present, however, the orbital/angular momentum transitions in the conduction and the valence bands take place at similar magnetic field values. Thus, no exciton angular momentum transition are found in Fig.~5(f), but the lowest bright and dark exciton energy levels approach each other with $B$, and eventually the two states cross each other at $B=39.1$ T. Yet, they almost appear as a doublet, with the energy difference not exceeding more than 0.16 meV in the range from 30 to 50 T.

For comparison, $l_{x}=0$ and $l_{x}=-1$ exciton energy levels in the fully opened quantum ring of the cross section shown in Fig.~1(b) are also displayed in Fig.\:5(b). For convenience, these levels are shifted down by 45 meV. One may notice that the exciton ground state is bright ($l_{x}=0$) in the whole range of $B$, which is similar to our previous finding for quantum rings of rectangular cross section\cite{Tadic2011}. Furthermore, the energy difference between the lowest optically active and dark state increases with $B$,  thus  no crossing between the two is observed at even much higher magnetic fields. Also, we show in Fig.\:5(e) $f_{zx}=-1$ and $f_{zx}=-2$ energy levels in the unstrained RLQD, which are increased by 125 meV. The exciton ground energies of both the fully opened quantum ring and the unstrained RLQD vary more slowly than in the case of the strained RLQD's. The smaller diamagnetic shift is due to smaller mean square of the in-plane electron-hole separation\cite{Godoy2006}, and therefore it demonstrates that the exciton polarization increases due to strain, as inferred from Figs.~2 and 3.

\begin{figure}
        \begin{center}
	\includegraphics[width=7.5cm]{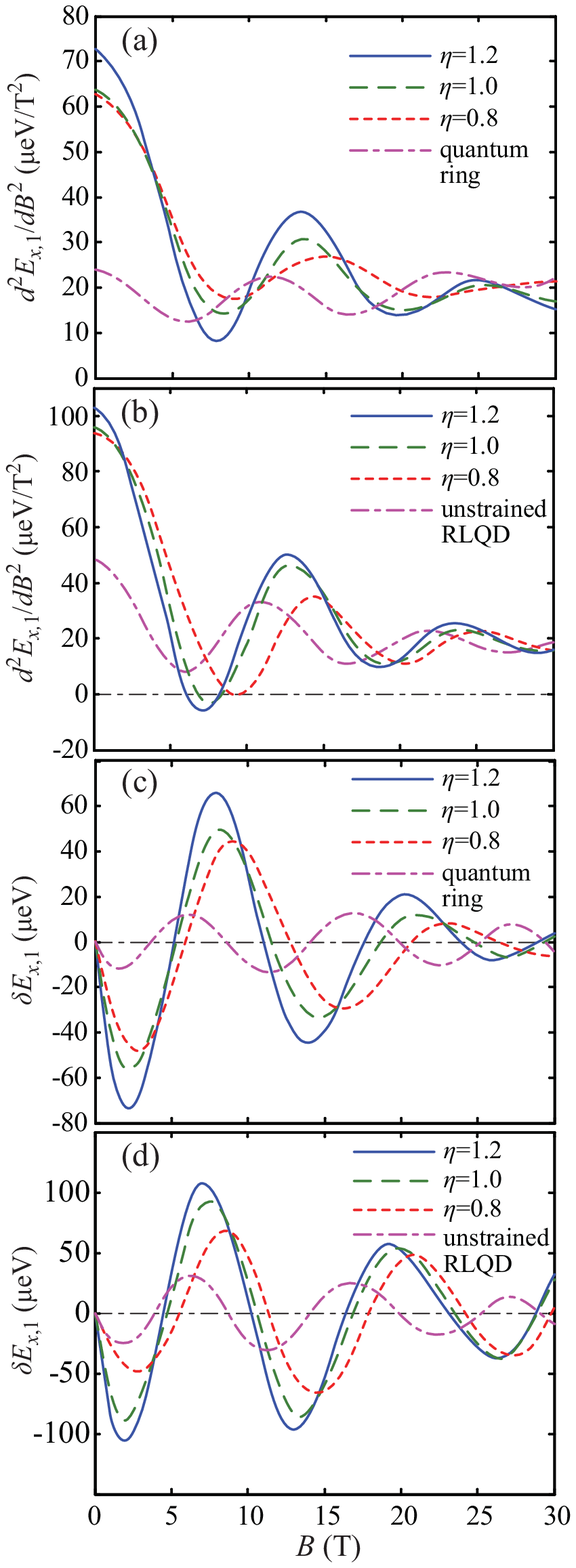}
	\caption{\label{fig6}(Color online.) {$d^2E_{x,1}/dB^2$} as
	function of {\it B} for: (a) the HH exciton (the dash-dotted {\color{mmagenta}
	magenta} line shows the result for the fully opened quantum ring), and
	(b) the multiband exciton (the dash-dotted {\color{mmagenta}
	magenta} line shows the result for the unstrained RLQD). Variation of the exciton energy residual
	{$\delta E_{x,1}$} with {\it B} for: (c) the HH exciton and (d) the multiband exciton.}
    \end{center}
\end{figure}

The oscillations of the lowest-energy bright exciton level with magnetic field
in Fig.\:5 are resolved by plotting {$d^2E_{x,1}/dB^2$} as a function of
{\it B} in Figs.\:6(a) and 6(b), for the cases of the HH and the multiband
exciton, respectively. Similar variations of the exciton energy residual
{$\delta E_{x,1}$}, defined in Eq.\:(34), with {\it B} are displayed in
Figs.\:6(c) and 6(d), for the HH and the multiband excitons, respectively.
One may notice that both {$d^2E_{x,1}/dB^2$} and {$\delta E_{x,1}$}
exhibit oscillations of decreasing magnitude when $\eta$ decreases. This is due to the peculiar strain distribution which leads to a barrier lowering inside the rim and raising inside the inner layer, as Fig.\:3(a) illustrates. It in turn  leads to a decrease of the exciton polarization.

The oscillations of the multiband exciton are found to be larger. In our
previous work we found that oscillations of the exciton ground energy level are established by means of anticrossings with higher exciton states of the
same orbital momentum\cite{Tadic2011}. These anticrossings appear to be more
effective when band mixing is taken into account. Nevertheless, the amplitudes of both the HH and multiband exciton energy level oscillations are of the same order of magnitude, which is 0.1 meV. It  agrees well with recent measurements on self-assembled type-I quantum rings\cite{Ding2010,Teodoro2010}, but it is larger than what we previously computed for cup-shaped
islands\cite{Arsoski2012}. Furthermore, because the magnitude and period of the oscillations for different $\eta$ are close to each other, and three values of $\eta$ correspond to three cross sections of the realistic structure, we deduce that the lack of axial symmetry would indeed have a small effect on the excitonic AB variations, as inferred in Sec. I. One might deduce that for the case of the structure in-plane anisotropy, as one observed in the experiment of Ref.~\onlinecite{Fomin2007}, the variations of both $d^2E_{x,1}/dB^2$ and $\delta E_{x,1}$ with $B$ are close to the average height of the ring, i.e., the case $\eta=1$. The quantitative estimation of the effects of the in-plane anisotropy of the shape of the structure is beyond the scope of the present analysis.

Also, for the geometry of the quantum ring, shown in Fig.\:1(b), the oscillations shown by the dash-dotted lines in Figs.\:6(a) and 6(c), demonstrate that the excitonic AB effect in the fully opened case is smaller than the oscillations in the strained RLQD. Larger oscillations in the latter are mainly a result of the increased exciton polarization as a consequence of the presence of the inner layer and the associated peculiar strain distribution. Moreover, for $\eta=1$, the unstrained RLQD exhibits smaller oscillations than the strained RLQD, as shown by the dash-dotted lines in Figs.\:6(b) and 6(d). Therefore, comparisons between the unstrained and strained cases in Figs.\:6(b) and 6(d) indicate that the presence of strain is not detrimental for the existence of the excitonic AB effect, but it leads to an almost double enlargement of this effect.

\begin{figure}
        \begin{center}
	\includegraphics[width=7cm]{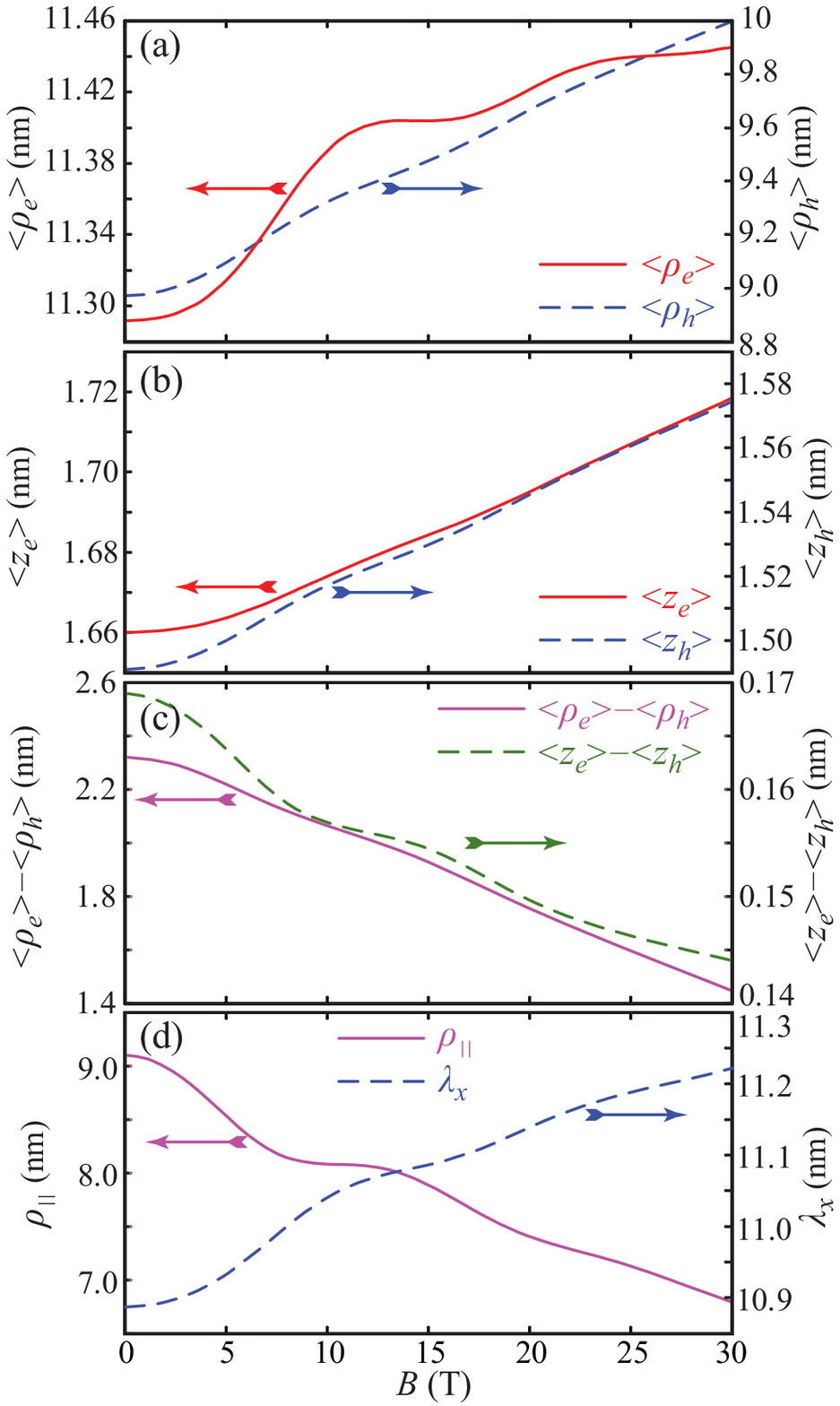}
	\caption{\label{fig7}(Color online) Variations of the parameters
	of the lowest-energy bright HH exciton state with {\it B} for {$\eta=1.0$}:
	(a) {$<\rho_{e}>$} (solid {\color{mred} red} line) and {$<\rho_{h}>$}
	(dashed {\color{mblue} blue} line) radius,
	(b) {$<z_{e}>$} (solid {\color{mred} red} line) and
	{$<z_{h}>$} (dashed {\color{mblue} blue} line), (c) {$<\rho_{e}>-<\rho_{h}>$}
	(solid {\color{mmagenta} magenta} line) and {$<z_{e}>-<z_{h}>$} (dashed {\color{mgreen} green} line),
	and (d) {$\rho_{\parallel}$} (solid {\color{mmagenta} magenta} line) and {$\lambda_{x}$}
	(dashed {\color{mblue} blue} line).}
    \end{center}
\end{figure}

To explain in more detail the origin and shape of the exciton
energy oscillations with magnetic field, the characteristic parameters of the
lowest-energy bright HH exciton state as function of {\it B} are displayed in
Fig.\:7 for {$\eta$}=1.0. We first note that variation of the average
electron radius {$<\rho_{e}>$} shown in Fig.\:7(a) is
much smaller than the variation of the average hole radius {$<\rho_{h}>$},
which could be ascribed to the smaller electron effective mass.
Variations of the vertical electron and hole positions with {\it B}
shown in Fig.\:7(b) are practically negligible, and also the average
vertical positions of the electron and hole are almost equal. However,
{$<\rho_{e}>-<\rho_{h}>$} decreases considerably with {\it B},
which is demonstrated in Fig.\:7(c). This decrease is a result of the
increasing Coulomb interaction when magnetic field increases.
Therefore, the exciton polarization decreases, and consequently
the excitonic AB oscillations abate, as demonstrated in Figs.\:5 and 6.
Similarly to {$<\rho_{e}>-<\rho_{h}>$}, the average in-plane exciton
radius {$\rho_{\parallel}$} exhibits oscillations around a decreasing
function of {\it B}, as shown in Fig.\:7(d). However, due to the increase
of {$<\rho_{e}>$} with {\it B} and the smaller electron effective mass,
the effective exciton radius {$\lambda_{x}$} is an increasing function
of {\it B}. Furthermore, it oscillates similar to {$\rho_{\parallel}$},
but its change is much smaller, just 0.3 nm when {\it B} increases
from 0 to 30 T. Also, in this magnetic field range the increase of
{$\lambda_{x}$} is smaller than the decrease of {$<\rho_{e}>-<\rho_{h}>$},
and thus one may deduce that the effective surface area is diminished.

\begin{figure}
        \begin{center}
	\includegraphics[width=7cm]{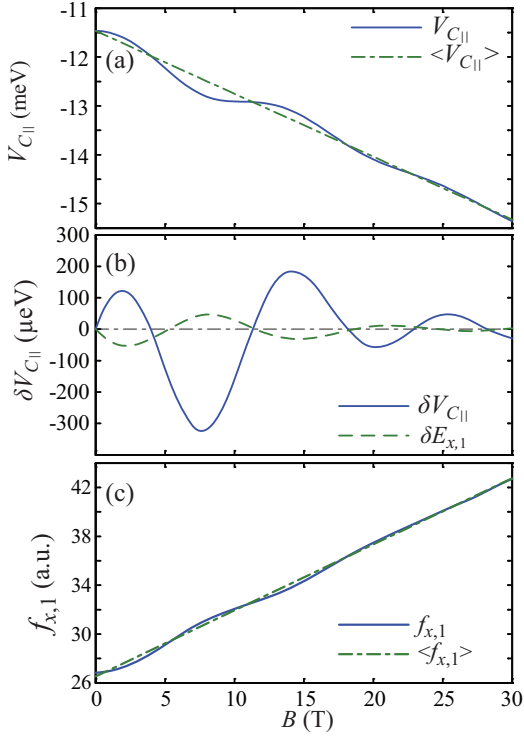}
           \caption{\label{fig8} (Color online) (a) The in-plane
	Coulomb potential {$V_{C\parallel}$} (solid {\color{mblue} blue} line)
	oscillates around {$\left<V_{C\parallel}\right>$} which is modeled
	by a linearly decreasing function of {\it B} (dashed {\color{mgreen} green} line).
	(b) Variation of the residuals of the in-plane Coulomb potential
	{$\delta V_{C \parallel}$} (solid {\color{mblue} blue} line) and the exciton energy
	{$\delta E_{x,1}$} (dashed {\color{mgreen} green} line) with magnetic field.
	(c) The oscillator strength for recombination of the lowest-energy
	bright exciton state (solid {\color{mblue} blue} line)
	exhibits oscillations around {$\left<f_{x,1}\right>$} the linear function of {\it B}
	(dashed {\color{mgreen} green} line).
	The results for the lowest-energy bright HH exciton state for {$\eta$}=1.0 are displayed.}
      \end{center}
\end{figure}

Because the RLQD's electron and hole states are better confined when {\it B} increases, the
overlap integrals between the single-particle states increase with {\it B}. It leads to the decrease
of $\rho_{\parallel}$ shown in Fig.\:7(d), and in turn $V_{C\parallel}$ decreases.  Also, it exhibits oscillations around
the linear function of $B$, as Fig.\:8(a) shows. By careful inspection of Figs.\:7(d) and 8(a), one
may find that the oscillations of $V_{C\parallel}$ are in phase with the oscillations of $\rho_{\parallel}$. The excitonic
Aharonov-Bohm oscillations are well resolved in Fig.\:8(b), where the plots of $\delta V_{C\parallel}(B)$ and
$\delta E_{x,1}(B)$ are both shown. These two quantities oscillate opposite to each other, illustrating
the fact that the Coulomb interaction tends to suppress the excitonic Aharonov-Bohm oscillations.
As a consequence, the oscillations of $\delta E_{x,1}(B)$ have a smaller amplitude than the
oscillations of $\delta V_{C\parallel}(B)$. Figure\:8(c) displays that opposite to the variation of $V_{C\parallel}$ with $B$,
enlarged confinement of the single-particle states causes the oscillator strength for recombination
of the lowest-energy bright exciton state $f_{x,1}$ with $B$ to increase.
Furthermore, $f_{x,1}$ exhibits oscillations around a linear function of {\it B} which are opposite to
the oscillations of $V_{C\parallel}$. Those oscillations  are nicely correlated with the oscillations of $\rho_{\parallel}$
shown in Fig.\:7(d). For example, whenever $\rho_{\parallel}$ exceeds the linear fitting function of $B$, $f_{x,1}$
drops below the similar linear dependence on $B$.

\begin{figure}
    \begin{center}
        \includegraphics[width=7cm]{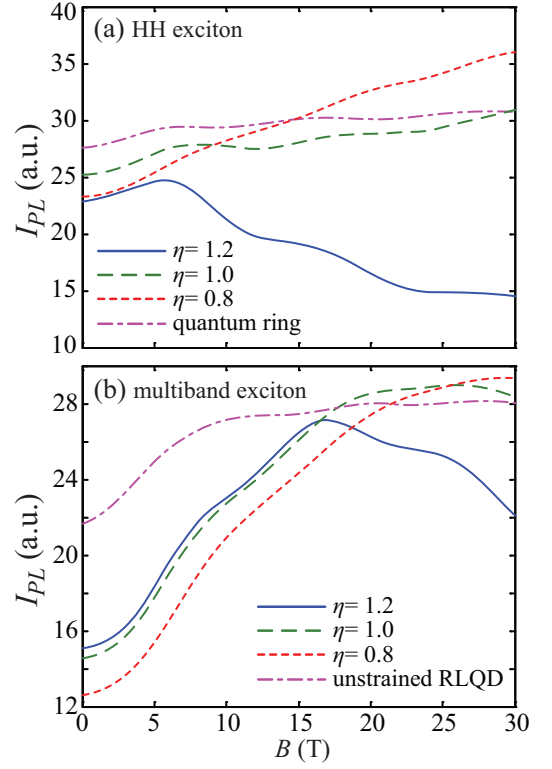}
	\caption{\label{fig9}(Color online) (a) Photoluminescence intensity
	{$I_{PL}$} for recombination of the HH exciton as function of {\it B}
	(the result for the fully opened quantum ring is shown by the
	dash-dotted {\color{mmagenta} magenta} line). (b) {$I_{PL}$}
	for recombination of the multiband exciton (the result for the unstrained
	RLQD is shown by the dash-dotted {\color{mmagenta} magenta} line).}
    \end{center}
\end{figure}

Figures\:9(a) and 9(b) show how the photoluminescence intensity varies with
{\it B} for the cases of the HH and multiband exciton, respectively. Because the exciton is dominantly localized in the bright exciton ground state
for small {\it B}, the {$I_{PL}$} dependence on {\it B} is similar to the
variation of {$f_{x,1}$} shown in Fig.\:8(c). However, {$I_{PL}$} decreases for $\eta=1.2$, because of a finite population of higher exciton dark states at finite temperature and the smaller energy difference with the bright exciton states. Also, all cases shown in Fig. 9, including the RLQD where strain is discarded, exhibit an oscillatory variation of $I_{PL}$ with $B$.

\section{Summary and conclusion}

We explored how the geometry affects the neutral exciton states in
quantum dots whose shape resembles rings but have a layer inside the
nominal ring opening. Our calculations reveal that such an inner layer
enhances the excitonic AB oscillations for the In(Ga)As/GaAs system,
and are larger than in the case of fully opened quantum rings. We show
that an increase of the height of the quantum-dot rim has the opposite
effect on the orbital (angular) momentum transitions, which is ascribed
to effects due to strain. As a matter of fact, we found that strain leads to
an increased separation between the electron and hole, and also it
reduces the mixing between the HH and LH states. The results of the
single band approximation are found to compare favorably well with the
multiband calculations, except that slightly larger oscillations of the
exciton energy levels are found if band mixing is taken into account.
Even a small increase of the rim height is found to bring about a
considerable shift of the angular momentum transition between the
exciton states. The magnetic field dependence of the photoluminescence
intensity is also affected by the presence of the inner layer and the
variation of the rim height. The magnitude of the computed excitonic AB
oscillations are found to be comparable to those measured
experimentally.

\section*{Acknowledgments}

This work was supported by the EU NoE: SANDiE, the Ministry of
Education, Science and Technological Development of Serbia, and the
Flemish Science Foundation (FWO-Vl).


\begin{thebibliography}{00}
\bibitem{Kastner2000}M. A. Kastner, Ann. Phys. {\bf 9}, 885 (2000).
\bibitem{Michler2000}P. Michler, A. Kiraz, C. Becher, W. V. Schoenfeld, P. M.
Petroff, L. Zhang, E. Hu, A. Imamogulu, Science {\bf 290}, 2282
(2000).
\bibitem{Engel2004}H. A. Engel, L. P. Kouwenhoven,
D. Loss, and C. M. Marcus, Quantum
Information Processing {\bf 3}, 115 (2004).
\bibitem{Engel2001}H. A. Engel, P. Recher, and D. Loss, Solid State Comm. {\bf 119}, 229 (2001).
\bibitem{Garcia1997}J. M. Garc\'ia, G. Medeiros-Ribeiro, K. Schmidt, T. Ngo, J. L.
Feng, A. Lorke, J. Kotthaus, and P. M. Petroff, Appl. Phys. Lett. {\bf 71}, 2014 (1997).
\bibitem{Teodoro2012}M. D. Teodoro, A. Malachias, V. Lopes-Oliveira, D. F. Cesar, V. Lopez-Richard, G. E. Marques, E. Marega, M. Benamara, Yu. I. Mazur, and G. J. Salamo, J. Appl. Phys. {\bf 112}, 014319 (2012).
\bibitem{Lee2004}B. C. Lee, O. Voskoboynikov, and C. P. Lee, Physica E {\bf 24}, 87 (2004).
\bibitem{AB1959}Y. Aharonov and D. Bohm, Phys. Rev. {\bf 115}, 485 (1959).
\bibitem{Govorov2002}A. O. Govorov, S. E. Ulloa, K. Karrai, and R. J. Warburton,
Phys. Rev. B {\bf 66}, 081309 (2002).
\bibitem{Ding2010}F. Ding, N. Akopian, B. Li, U. Perinetti, A. Govorov, F. M. Peeters,
C. C. Bof Bufon, C. Deneke, Y. H. Chen, A. Rastelli, O. G. Schmidt, and V. Zwiller,
Phys. Rev. B {\bf82}, 075309 (2010).
\bibitem{Teodoro2010}M. D. Teodoro, V. L. Campo, Jr., V. Lopez-Richard, E. Marega,
Jr., G. E. Marques, Y. G. Gobato, F. Iikawa, M. J. S. P. Brasil, Z. Y. Abu Waar,
V. G. Dorogan, Yu. I. Mazur, M. Benamara, and G. J. Salamo,
Phys. Rev. Lett. {\bf 104}, 086401 (2010).
\bibitem{Fomin2007}V. M. Fomin, V. N. Gladilin, S. N. Klimin, J. T. Devreese, N.
A. J. M. Kleemans, and P. M. Koenraad, Phys. Rev. B {\bf 76}, 235320 (2007).
\bibitem{Lorke2002}A. Lorke, R. Blossey, J. M. Garc\'ia, M. Bichler, and G. Abstreiter,
Mater. Sci. Eng. B {\bf 88}, 225 (2002).
\bibitem{Granados2003}D. Granados and J. M. Garc\'ia, Appl. Phys. Lett. {\bf 82}, 2401 (2003).
\bibitem{Kleemans2007}N. A. J. M. Kleemans, I. M. A. Bominaar-Silkens, V. M. Fomin,
V. N. Gladilin, D. Granados, A. G. Taboada, J. M. Garc\'ia, P. Offermans,
U. Zeitler, P. C. M. Christianen, J. C. Maan, J. T. Devreese, and P. M.
Koenraad, Phys. Rev. Lett. {\bf 99}, 146808 (2007).
\bibitem{Offermans2005}P. Offermans, P. M. Koenraad, J. H. Wolter, D. Granados, J. M.
Garc\'ia, V. M. Fomin, N. Gladilin, and J. T. Devreese, Appl. Phys. Lett. {\bf 87}, 131902 (2005).
\bibitem{Arsoski2012}V. Arsoski, N. \v{C}ukari\'c, M. Tadi\'c, and F. M. Peeters, Phys. Scr.
{\bf 2012}, 014054 (2012).
\bibitem{Li2011} B.Li and F.M. Peeters, Phys. Rev. B {\bf 83}, 115448 (2011).
\bibitem{Cukaric2012}N. \v{C}ukari\'c, V. Arsoski, M. Tadi\'c, and F.M. Peeters, Phys. Rev. B
{\bf 85}, 235425 (2012).
\bibitem{Tadic2011}M. Tadi\'c, N. \v{C}ukari\'c, V. Arsoski, and F.M. Peeters, Phys. Rev.
B {\bf 84}, 125307 (2011).
\bibitem{Filikhin2006}I. Filikhin, V. M. Suslov and B. Vlahovic, Physica E {\bf 33}, 349 (2006).
\bibitem{Climente2003}J. I. Climente, J. Planelles, and W. Jasko\'lski, Phys. Rev. B
{\bf 68}, 075307 (2003).
\bibitem{Ribeiro2004}E. Ribeiro, A. O. Govorov, W. Carvalho Jr., and G.
Medeiros-Ribeiro, Phys. Rev. Lett. {\bf 92}, 126402 (2004).
\bibitem{Davies1998}J. H. Davies, J. Appl. Phys. {\bf 84}, 1358 (1998).
\bibitem{Tadic2002}M. Tadi\'c, F. M. Peeters, K. L. Janssens, M. Korkusi\'nski, and P. Hawrylak,
J. Appl. Phys. {\bf 92}, 5819 (2002).
\bibitem{Eshelby1957}J.D. Eshelby, Proc. R. Soc. London Ser. A {\bf 241}, 376 (1957).
\bibitem{Downes1997}J. R. Downes, D. A. Faux, and E. P. O'Reilly, J. Appl. Phys.
{\bf81}, 6700 (1997).
\bibitem{Chuang1991}S.L. Chuang, Phys. Rev. B {\bf 43}, 9649 (1991).
\bibitem{Pedersen1996}F. B. Pedersen and Y. C. Chang, Phys. Rev. B {\bf 53},
1507 (1996).
\bibitem{TadicPRB2002}M. Tadi\'c, F. M. Peeters, and K. L. Janssens,
Phys. Rev. B {\bf 65}, 165333 (2002).
\bibitem{Barticevic2006}Z. Barticevic, M. Pacheco, J. Simonin, and C.R. Proetto, Phys. Rev. B {\bf 73}, 165311 (2006).
\bibitem{Grochol2006}M. Grochol, F. Grosse, and R. Zimmermann, Phys. Rev. B
{\bf 74}, 115416 (2006).
\bibitem{Efros1996}Al. L. Efros, M. Rosen, M. Kuno, M. Nirmal, D. J. Norris, and
M. Bawendi, Phys. Rev. B {\bf 54}, 4843 (1996).
\bibitem{Degani2008}M. H. Degani, M. Z. Maialle, G. Medeiros-Ribeiro, and E.
Ribeiro, Phys. Rev. B {\bf 78}, 075322 (2008).
\bibitem{Vurgaftman2001}I. Vurgaftman, J. R. Meyer, and L. R. Ram-Mohan, J. Appl.
Phys. {\bf 89}, 5815 (2001).
\bibitem{Tanaka1999}K. Tanaka, N. Kotera, and H. Nakamura, J. Appl. Phys. {\bf 85}, 4071 (1999).
\bibitem{Alavi1980}K. Alavi, R. L. Aggarwal, and S. H. Groves, Phys. Rev. B {\bf 21}, 1311 (1980).
\bibitem{Lawaetz1971}P. Lawaetz, Phys. Rev. B {\bf 4}, 3460 (1971).
\bibitem{Bayer2004}M. Bayer, A. Kuther, A. Forchel, A. Gorbunov, V. B. Timofeev,
F. Sch\"afer, J. P. Reithmaier, T. L. Reinecke, and S. N.Walck,
Phys. Rev. Lett. {\bf 82}, 1748 (1999).
\bibitem{Nakaoka2004}T. Nakaoka, T. Saito, J. Tatebayashi, and Y. Arakawa, Phys. Rev. B {\bf 70}, 235337 (2004).
\bibitem{Kleemans2009}N. A. J. M. Kleemans, J. van Bree, M. Bozkurt, P. J. van Veldhoven,
P. A. Nouwens, R. N\"otzel, A. Yu. Silov, P. M. Koenraad, and M. E. Flatt\'e, Phys. Rev. B
{\bf 79}, 045311 (2009).
\bibitem{Pryor2006}C. E. Pryor and M. E. Flatt\'e, Phys. Rev. Lett. {\bf 96}, 026804 (2006);
see erratum {\bf 99}, 179901(E) (2007).
\bibitem{Godoy2006}M. P. F. de Godoy, P. F. Gomes, M. K. K. Nakaema, F. Iikawa, M. J. S. P. Brasil, R. A. Caetano, J. R. Madureira, J. R. R. Bortoleto, M. A. Cotta, E. Ribeiro, G. E. Marques, and A. C. R. Bittencourt, Phys. Rev. B {\bf 73}, 033309 (2006).
\end{thebibliography}
\end{document}